\documentclass[11pt,twoside]{book} 
\usepackage{asp2008n}
\usepackage{times}
\usepackage{lscape}
\usepackage{epsf}

\usepackage{graphicx}
\usepackage{amssymb}
\usepackage{longtable}
\usepackage[figuresright]{rotating}
\usepackage{multirow}

\newcommand{\simless}{\mathbin{\lower 3pt\hbox {$\rlap{\raise 5pt\hbox{$\char'074$}}\mathchar"7218$}}}

\mainmatter

\newlength{\deftabcolsep}
\begin{document}


\title{The Young Cluster and Star Forming Region NGC\,2264}   
\author{S. E. Dahm}   
\affil{Department of Astronomy, California Institute of Technology, MS 105-24, Pasadena, CA 91125, USA}    

\begin{abstract} 
NGC\,2264 is a young Galactic cluster and the dominant component of the Mon OB1 association
lying approximately 760 pc distant within the local spiral arm. The cluster is hierarchically
structured, with subclusters of suspected members spread across several parsecs.
Associated with the cluster is an extensive molecular cloud complex spanning more than two
degrees on the sky. The combined mass of the remaining molecular cloud cores upon which the
cluster is superposed is estimated to be at least $\sim$3.7$\times$10$^{4}$ M$_{\odot}$.
Star formation is ongoing within the region as evidenced by the presence of numerous embedded
clusters of protostars, molecular outflows, and Herbig-Haro objects. The stellar population
of NGC\,2264 is dominated by the O7 V multiple star, S~Mon, and several dozen B-type zero-age
main sequence stars. X-ray imaging surveys, H$\alpha$ emission surveys, and photometric
variability studies have identified more than 600 intermediate and low-mass members
distributed throughout the molecular cloud complex, but concentrated within two densely
populated areas between S~Mon and the Cone Nebula. Estimates for the total stellar
population of the cluster range as high as 1000 members and limited deep photometric surveys
have identified $\sim$230 substellar mass candidates. The median age of NGC\,2264 is estimated
to be $\sim$3 Myr by fitting various pre-main sequence isochrones to the low-mass stellar
population, but an apparent age dispersion of at least $\sim$5 Myr can be inferred from the
broadened sequence of suspected members. Infrared and millimeter observations of the cluster
have identified two prominent sites of star formation activity centered near NGC\,2264 IRS1,
a deeply embedded early-type (B2--B5) star, and IRS2, a star forming core and associated
protostellar cluster. NGC\,2264 and its associated molecular clouds have been extensively
examined at all wavelengths, from the centimeter regime to X-rays. Given its relative
proximity, well-defined stellar population, and low foreground extinction, the cluster
will remain a prime candidate for star formation studies throughout the foreseeable future.
\end{abstract}

\section{Introduction}

Few star forming regions surpass the resplendent beauty of NGC\,2264, the richly populated
Galactic cluster in the Mon OB1 association lying approximately 760 pc distant in the local
spiral arm. Other than the Orion Nebula Cluster, no other star forming region within
one kpc possesses such a broad mass spectrum and well-defined pre-main sequence population
within a relatively confined region on the sky. Estimates for the total stellar population
of the cluster range up to $\sim$1000 members, with most low-mass, pre-main sequence stars having
been identified from H$\alpha$ emission surveys, X-ray observations by {\it ROSAT}, {\it Chandra},
and {\it XMM-Newton}, or by photometric variability programs that have found several hundred
periodic and irregular variables. The cluster of stars is seen in projection against an extensive
molecular cloud complex spanning more than two degrees north and west of the cluster center.
The faint glow of Balmer line
emission induced by the ionizing flux of the cluster OB stellar population contrasts starkly
with the background dark molecular cloud from which the cluster has emerged. The dominant stellar
member of NGC\,2264 is the O7 V star, S~Monocerotis (S~Mon), a massive multiple star lying in the
northern half of the cluster. Approximately 40\arcmin\ south of S~Mon is the prominent Cone Nebula,
a triangular projection of molecular gas illuminated by S~Mon and the early B-type cluster members.
NGC\,2264 is exceptionally well-studied at all wavelengths: in the millimeter by Crutcher et al. (1978),
Margulis \& Lada (1986), Oliver et al. (1996), and Peretto et al. (2006); in the
near infrared (NIR) by Allen (1972), Pich\'{e} (1992, 1993), Lada et al. (1993), Rebull et al. (2002),
and Young et al. (2006); in the optical by Walker (1956), Rydgren (1977), Mendoza \& G\'omez (1980),
Adams et al. (1983), Sagar \& Joshi (1983), Sung et al. (1997), Flaccomio et al. (1999),
Rebull et al. (2002), Sung et al. (2004), Lamm et al. (2004), and Dahm \& Simon (2005); and in X-rays by Flaccomio et al. (2000), Ramirez et al. (2004),
Rebull et al. (2006), Flaccomio et al. (2006), and Dahm et al. (2007).

NGC\,2264 was discovered by Friedrich Wilhelm Herschel in 1784 and listed as H VIII.5 in his
catalog of nebulae and stellar clusters. The nebulosity associated with NGC\,2264 was also
observed by Herschel nearly two years later and assigned the designation: H V.27. The Roman
numerals in Herschel's catalog are object identifiers, with `V' referring to very large nebulae
and `VIII' to coarsely scattered clusters of stars. One of the first appearances of the cluster
in professional astronomical journals is Wolf's (1924) reproduction of a photographic plate of
the cluster and a list of 20 suspected variables.
Modern investigations of the cluster begin with Herbig (1954) who used the slitless
grating spectrograph on the Crossley reflector at Lick Observatory to identify 84 H$\alpha$ emission
stars, predominantly T Tauri stars (TTS), in the cluster region. Herbig (1954) postulated that these
stars represented a young stellar population emerging from the dark nebula. Walker's (1956) seminal
photometric and spectroscopic study of NGC\,2264 discovered that a normal main sequence exists from
approximately O7 to A0, but that lower mass stars consistently fall above the main sequence.
This observation was in agreement with predictions of early models
of gravitational collapse by Salpeter and by Henyey et al. (1955). Walker (1954, 1956) proposed
that these stars represent an extremely young population of cluster members, still undergoing
gravitational contraction. Walker (1956) further noted that the TTSs within the cluster fall above
the main sequence and that they too may be undergoing gravitational collapse. Walker (1956) concluded
that the study of TTSs would be ``of great importance for our understanding of these early stages of
stellar evolution.''

\begin{figure}[!tbh]
\begin{center}
\includegraphics[angle=90,width=\linewidth, draft=False]{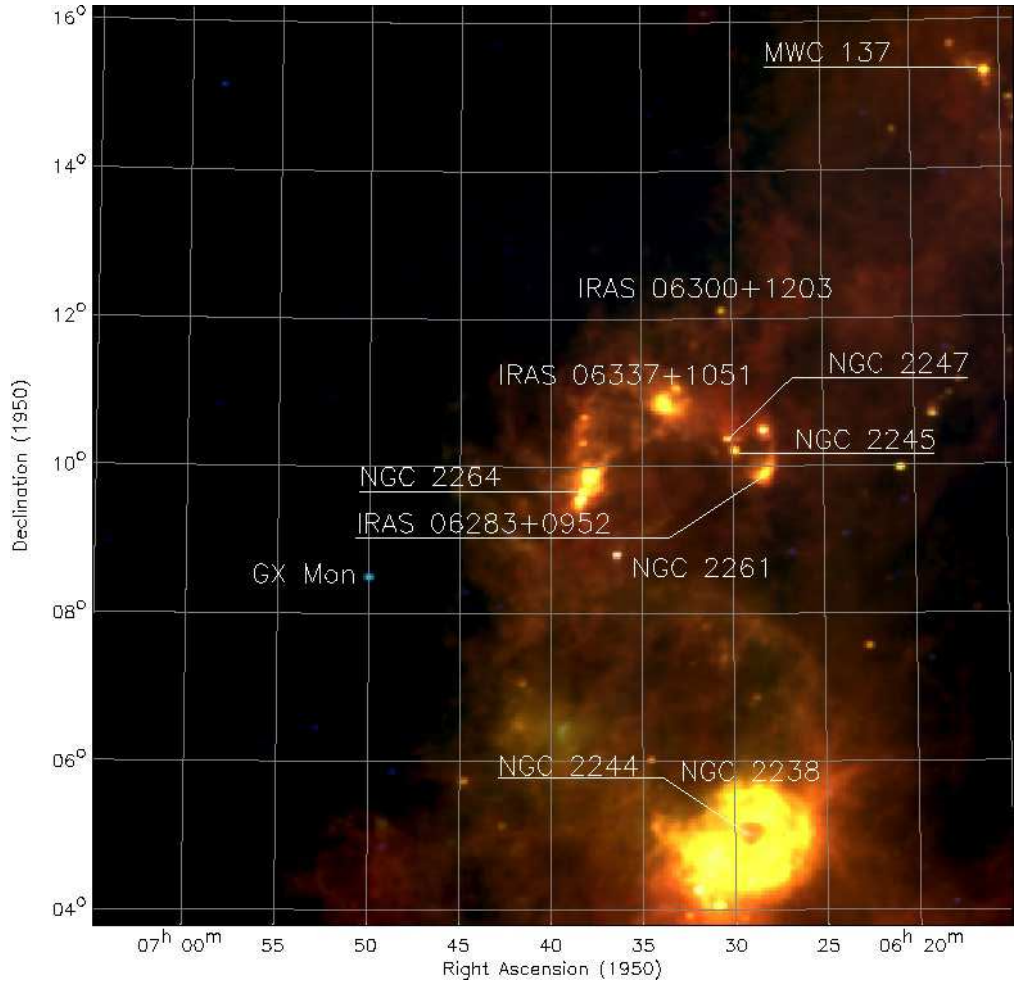}
\caption[fg1.eps]{A 12.5$^{\circ}$ square false-color IRAS image (100, 60 \& 25~$\mu$m) of the Mon OB1 and Mon R1
associations. NGC\,2264 lies at the center of the image with several nearby IRAS sources identified, including
the reflection nebulae NGC\,2245 and NGC\,2247, and NGC\,2261 (Hubble's Variable Nebula). South of NGC\,2264
is the Rosette Nebula and its embedded cluster NGC\,2244, lying 1.7 kpc distant in the Perseus arm. \label{f1}}
\end{center}
\end{figure}

The molecular cloud complex associated with NGC\,2264 was found by Crutcher et al. (1978) to consist
of several cloud cores, the most massive of which lies roughly between S~Mon and the Cone Nebula.
Throughout the entire cluster region, Oliver et al. (1996) identified 20 molecular clouds ranging
in mass from $\sim10^{2}$ to $10^{4}$ M$_{\odot}$. With NGC\,2264 these molecular clouds comprise
what is generally regarded as the Mon OB 1 association. Active star formation is ongoing within NGC\,2264
as evidenced by the presence of numerous embedded protostars and clusters of stars, as well as molecular
outflows and Herbig-Haro objects (Adams et al. 1979; Fukui 1989; Hodapp 1994; Walsh et al. 1992; Reipurth et al. 2004a;
Young et al. 2006). Two prominent sites of star formation activity within the cluster are IRS1 (also known as Allen's source), located
several arcminutes north of the tip of the Cone Nebula, and IRS2, which lies approximately one-third of
the distance from the Cone Nebula to S~Mon. New star formation activity is also suspected within the
northern extension of the molecular cloud based upon the presence of several embedded IRAS sources and giant Herbig-Haro flows (Reipurth et al. 2004a,c).
From 60 and 100~$\mu$m IRAS images of NGC\,2264, Schwartz (1987) found that the cluster lies on the eastern edge
of a ring-like dust structure, 3$^{\circ}$ in diameter. Shown in Figure 1 is a 12\fdg5$\times$12\fdg5 false-color
IRAS image (100, 60, and 25~$\mu$m) centered near NGC\,2264. The reflection nebulae NGC\,2245 and NGC\,2247,
members of the Mon R1 association, are on the western boundary of this ring (see the chapter by Carpenter \& Hodapp).
Other components of the Mon R1 association include the reflection nebulae IC\,446 and IC\,2169, LkH$\alpha$215,
as well as several early type (B3--B7) stars. It is generally believed that the Mon R1 and Mon OB 1 associations
are at similar distances and are likely related. The Rosette Nebula, NGC\,2237-9, and its embedded
young cluster NGC\,2244 lie 5$^{\circ}$ southwest of NGC\,2264, 1.7 kpc distant in the outer
Perseus arm (see the chapter by Rom\'an-Z\'u\~niga \& Lada). Several arcs of dust and CO emission have been
identified in the region, which are believed to be supernovae remnants or windblown shells. Many of these
features are apparent in Figure 2, a wide-field H$\alpha$ image of NGC\,2264, NGC\,2244, and the intervening
region obtained by T. Hallas and reproduced here with his permission. It is possible that star formation
in the Mon OB1 and R1 associations was triggered by nearby energetic events, but it is difficult to assess
the radial distance of the ringlike structures evident in Figure 2, which may lie within the Perseus arm or
the interarm region. Shown in Figure 3 is a narrow-band composite image of NGC\,2264 obtained by T.A. Rector
and B.A. Wolpa using the 0.9 meter telescope at Kitt Peak. S~Mon dominates the northern half of the cluster,
which lies embedded within the extensive molecular cloud complex.

\begin{figure}
\centering
\includegraphics[angle=0,width=\linewidth, draft=False]{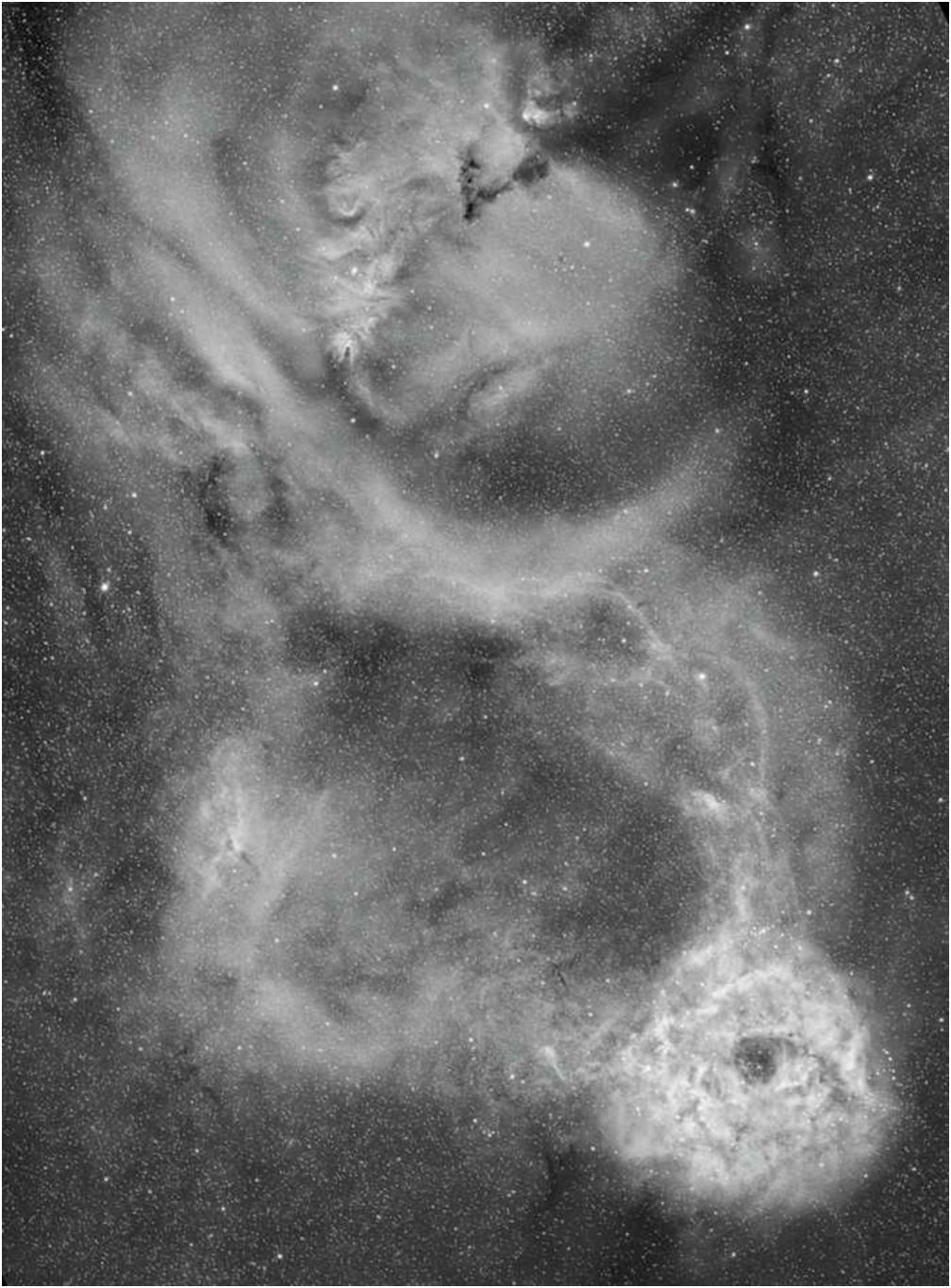}
\caption[fg2.eps]{An extraordinary widefield, narrow-band H$\alpha$ image of NGC\,2264 (upper center), the Rosette
Nebula and NGC\,2244 (lower right), and the numerous windblown shells and supernova remnants possibly associated
with the Mon OB1 or Mon OB2 associations. The Cone Nebula is readily visible just above and left of image center
as is S~Mon. Also apparent in the image is the dark molecular cloud complex lying to the west of NGC\,2264.
This image is a composite of 16 20-minute integrations obtained by T. Hallas using a 165 mm lens and an Astrodon
H$\alpha$ filter.
\label{f2}}
\end{figure}

To summarize all work completed over the last half-century in NGC\,2264  would be an overwhelming task and require
significantly more pages than alloted for this review chapter. The literature database for NGC\,2264 and its members
has now grown to over 400 refereed journal articles, conference proceedings, or abstracts. Here we attempt to highlight
large surveys of the cluster at all wavelengths as well as bring attention to more focused studies of the cluster that
have broadly impacted our understanding of star formation. The chapter begins with a review of basic cluster properties
including distance, reddening, age, and inferred age dispersion. It then examines the OB stellar population of the
cluster, the intermediate and low-mass stars, and finally the substellar mass regime. Different wavelength
regions are examined from the centimeter, millimeter, and submillimeter to the far-, mid-, and near infrared,
the optical, and the X-ray regimes. We then review many photometric variability studies of the cluster that have
identified several hundred candidate members. Finally, we consider future observations of the cluster and what additional
science remains to be reaped from NGC\,2264. The cluster has remained in the spotlight of star formation studies for more
than 50 years, beginning with the H$\alpha$ survey of Herbig (1954). Its relative proximity, low foreground
extinction, large main sequence and pre-main sequence populations, the lack of intense nebular emission, and the
tremendous available archive of observations of the cluster at all wavelengths guarantee its place with the
Orion Nebula Cluster and the Taurus-Auriga molecular clouds as the most accessible and observed Galactic star forming region.

\begin{figure}[!tbh]
\vspace{8mm}
\begin{center}
\includegraphics[angle=0,width=\linewidth, draft=False]{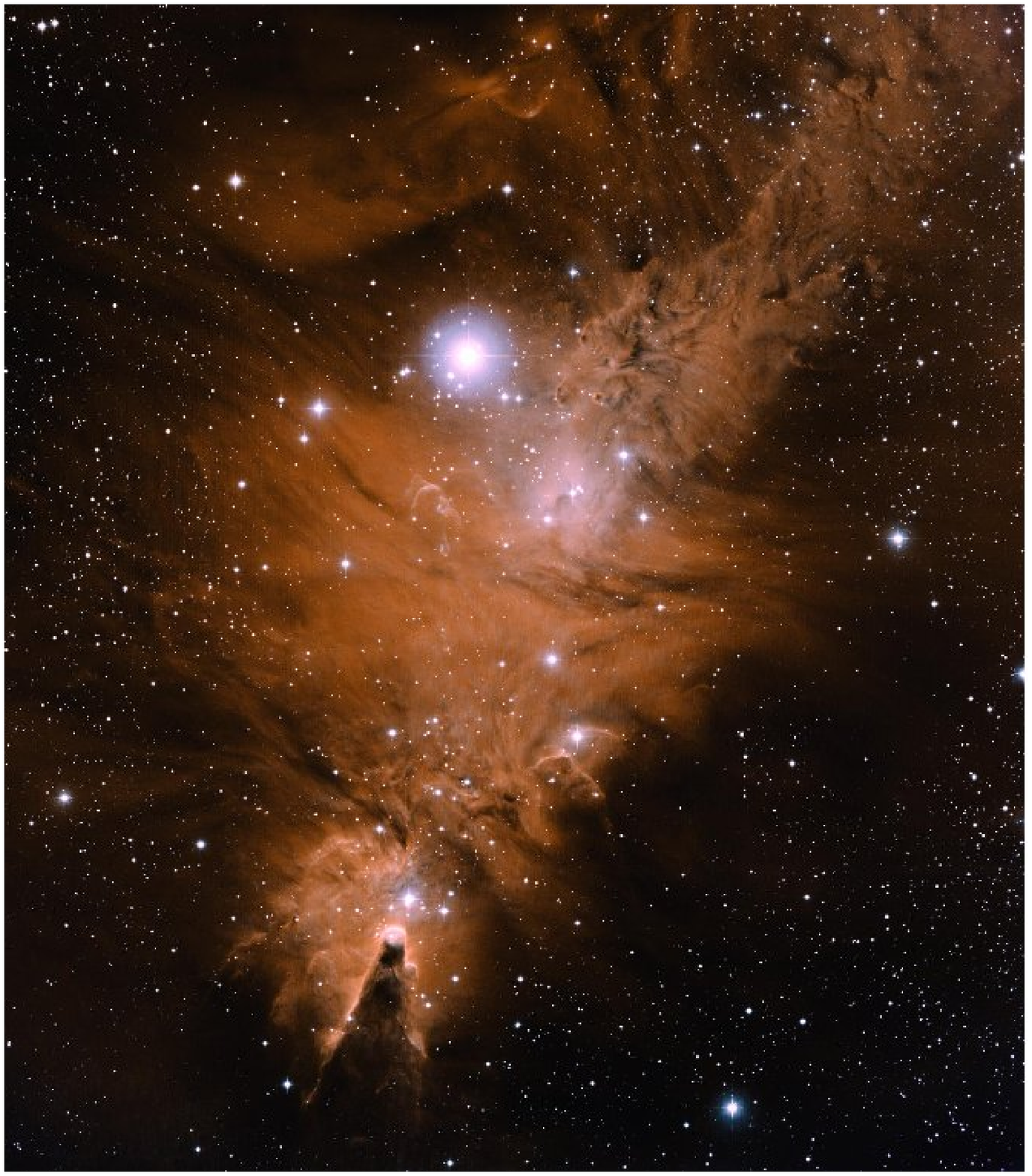}
\caption[fg3.eps]{Narrow-band, three-color image of NGC\,2264 obtained by T.A. Rector and B.A. Wolpa (NOAO/AURA/NSF)
using the 0.9 meter telescope at Kitt Peak National Observatory. The filters used for this composite image are: O III (light blue),
H$\alpha$ (red-orange), and [S II] (blue-violet). The field of view is approximately 0.75$^{\circ}$$\times$1$^{\circ}$. S~Mon lies
just above the image center and is believed to be the ionizing source of the bright rimmed Cone Nebula. \label{f2}}
\end{center}
\vspace{-8mm}
\end{figure}

\section{Cluster Distance and Interstellar Extinction}

 NGC\,2264 is ideally suited for accurate distance determinations given its lack of significant
foreground extinction and the abundant numbers of early-type members. Difficulty in establishing
the cluster distance, however, arises from the near vertical slope of the zero age main sequence
(ZAMS) for OB stars in the color-magnitude diagram and from the depth of the cluster along the
line of sight, which has not been assessed.  The distance of NGC\,2264 is now widely accepted to
be $\sim$760 pc, but estimates found in the early literature vary significantly. Walker (1954) used
photoelectric observations of suspected cluster members earlier than A0 to derive a distance modulus
of 10.4 mag (1200 pc) by fitting the standard main sequence of Johnson \& Morgan (1953). Herbig (1954)
adopted a distance of 700 pc, the mean of published values available at the time. For his landmark study
of the cluster, Walker (1956) revised his earlier distance estimate to 800 pc using the modified main
sequence of Johnson \& Hiltner (1956). P\'{e}rez et al. (1987) redetermined the distances
of NGC\,2264 and NGC\,2244 (the Rosette Nebula cluster) assuming an anomalous ratio of total-to-selective
absorption. Their revised distance estimate for NGC\,2264 was $950 \pm 75$ pc. Included in their study
is an excellent summary of distance determinations for NGC\,2244 and NGC\,2264 found in the literature
from 1950 to 1985 (their Table XI). The mean of these values for NGC\,2264 is $735 \pm 106$ pc,
significantly less than their adopted distance. P\'{e}rez (1991) provides various physical characteristics
of the cluster including distance, total mass, radius, mean radial velocity, and age, but later
investigations of the cluster have since revised many of these estimates. From modern CCD photometry
of NGC\,2264, Sung et al. (1997), determined
a mean distance modulus of $V_0 - M_V = 9.4 \pm 0.25$ or 760 pc using 13 B-type cluster members, which have
distance moduli in the range from $9-10$ mag. This value is now cited in most investigations of the cluster.
The projected linear dimension of the giant molecular cloud associated with NGC\,2264 including the northern
extension is nearly 28 pc. If a similar depth is assumed along the cluster line of sight, an intrinsic
uncertainty of nearly 4\% is introduced into the distance determination.

Interstellar reddening toward NGC\,2264 is recognized to be quite low. Walker (1956)
found $E(B-V) = 0.082$ or $A_V = 0.25$ assuming the normal ratio of total-to-selective absorption, $R = 3.08$.
P\'{e}rez et al. (1987) found a similar reddening value with $E(B-V) = 0.061$, but used
$R = 3.63$ to derive their significantly greater distance to the cluster. In their comprehensive photometric study, Sung et al. (1997)
found the mean reddening of 21 OB stars within the cluster to be $E(B-V) = 0.071 \pm 0.033$,
in close agreement with the estimates of Walker (1956) and P\'{e}rez et al. (1987). No
significant deviation from these values has been found using the early-type cluster members. Individual
extinctions for the suspected low-mass members, however, are noted to be somewhat higher. Rebull et al. (2002)
derive a mean $E(B-V) = 0.146 \pm 0.03$ or $A_V = 0.45$ from their spectroscopic sample of more than 400 stars,
only 22\% of which are earlier than K0. Dahm \& Simon (2005) find that for the H$\alpha$ emitters within the
cluster with established spectral types, a mean A$_{V}$ of 0.71 mag follows from $A_{V} = 2.43 E(V-I_{c})$.
Some of these low-mass stars are suffering from local extinction effects (e.g. circumstellar disks) or lie
within deeply embedded regions of the molecular cloud. The mean extinction value derived for the OB stellar
population, which presumably lies on the main sequence and possesses well-established intrinsic colors,
should better represent the distance-induced interstellar reddening suffered by cluster members.
Table 1 summarizes the distance and extinction estimates of the cluster found or adopted in selected surveys of
NGC\,2264.

\begin{sidewaystable}
\begin{footnotesize}
\begin{center}
\begin{tabular}[c]{lcccccc}
\multicolumn{7}{c}{Table 1} \\
\multicolumn{7}{c}{Summary of Earlier Surveys of NGC\,2264} \\
\tableline
Authors                       & Age (Myr) & M$_{V}$ Range & Isochrone\tablenotemark{a} & E($B-V$) & Distance (pc)\tablenotemark{b} & Notes\tablenotemark{c}\\
\tableline
Walker (1956)                & 3.0      &  $-5.0 \le M_{V} \le +8.0$  & Henyey et al. (1955) & 0.082 & 800 & pe and pg\\
                             &          &                             &                      &       &     &          \\
Mendoza \& G\'{o}mez (1980)  & 3.0      &  $-5.0 \le M_{V} \le +4.0$  & Iben \& Talbot (1966)& 0.06  & 875 & pe       \\
                             &          &                             &                      &       &     &          \\
Adams et al. (1983)          & 3.0-6.0  &  $+1.0 \le M_{V} \le +11.0$ & Cohen \& Kuhi (1979) & [0.06] & [800] & pg\\
                             &          &                             &                      &       &     &          \\
Sagar \& Joshi (1983)        & 5        &  $-5.2 \le M_{V} \le +5.7$  & Cohen \& Kuhi (1979) & $\leq$0.12 & 794 & pe\\
                             &          &                             &                      &       &     &          \\
P\'{e}rez et al. (1987)      & ...      &  $-2.0 \le M_{V} \le +3.0$  & ...                  & 0.06  & 950 & pe\\
                             &          &                             &                      &       &     &          \\
Feldbrugge \& van Genderen (1991) & $< 3.0$ & $M_{V} \le +3.0$        & ...                  & 0.04  & 700 & pe\\
                             &          &                             &                      &       &     &          \\
Neri et al. (1993)	     & ...   	&  $-5.0 \le M_{V} \le +6.0$  & ... 		     & 0.05  & 910 & pe\\
                             &          &                             &                      &       &     &          \\
Sung et al. (1997)           & 0.8--8.0 &  $-2.0 \le M_{V} \le +7.0$  & S94, BM96            & 0.071 & 760 & CCD\\
                             &          &                             &                      &       &     &          \\
Flaccomio et al. (1999)      & 0.1--10.0&  $-0.3 \le M_{V} \le +7.2$  & DM97                 & [0.06]  & [760-950] & CCD \\
                             &          &                             &                      &       &     &          \\
Park et al. (2000)           & 0.9--4.3 &  $-5.0 \le M_{V} \le +7.0$  & DM94, S94, B98       & 0.066 & 760 & CCD \\
                             &          &                             &                      &       &     &          \\
Rebull et al. (2002)         & 0.1--6.0 &  $+3.0 \le M_{V} \le +8.5$  & DM94, SDF00          & 0.15  & [760] & CCD \\
                             &          &                             &                      &       &     &          \\
Sung et al. (2004)           & 3.1      &  $-2.0 \le M_{V} \le +10.0$ & SDF00                & 0.07--0.15 & ... & CCD\\
\tableline
\noalign{\smallskip}
\multicolumn{7}{l}{$^{a}$ S94 - Swenson et al. (1994); BM96 - Bernasconi \& Maeder (1996); DM94 - D'Antona \& Mazzitelli (1994); B98 - Baraffe et al. (1998);}\\
\multicolumn{7}{l}{~~~SDF00 - Siess et al. (2000)}\\
\multicolumn{7}{l}{$^{b}$ Brackets indicate adopted values (not determined).}\\
\multicolumn{7}{l}{$^{c}$ pe - photoelectric; pg - photographic}\\
\end{tabular}\\
\end{center}
\end{footnotesize}
\end{sidewaystable}

\section {The Age and Age Dispersion of NGC\,2264}

 The age of NGC\,2264 has long been inferred to be young given the large OB stellar population of the cluster
and the short main sequence lifetimes of these massive stars. Walker (1956) derived an estimate of 3 Myr, based upon
the main sequence contraction time of an A0-type star (the latest spectral type believed to be on the ZAMS)
from the theoretical work of Henyey et al. (1955). Iben \& Talbot (1966), however, directly compared theoretical
time-constant loci or isochrones with the color-magnitude diagram of NGC\,2264, concluding that star formation began within
the cluster more than 65 Myr ago. They further suggested that the star formation rate has been increasing exponentially with time,
and that the average mass of each subsequent generation of stars has also increased exponentially. Strom et al. (1971)
and Strom et al. (1972), however, in their insightful investigations of Balmer line emission and infrared excesses
among A and F-type members of NGC\,2264, concluded that dust and gas shells (spherical or possibly disk-like) were
common among pre-main sequence stars, making strict interpretation of age dispersions from color-magnitude
diagrams difficult if not impossible. Strom et al. (1972) conclude that an intrinsic age dispersion of 1 to 3 Myr
is supported by the A and F-type members of NGC\,2264. Adams et al. (1983) revisited the question of age
dispersion within NGC\,2264 in their deep photometric survey of the cluster. From the theoretical HR diagram of
probable cluster members, a significant age spread of more than 10 Myr is inferred. Adams et al. (1983)
further suggest that sequential star formation has occurred within the cluster, beginning with the low-mass stars
and continuing with the formation of more massive cluster members. From the theoretical models of Cohen \& Kuhi (1979),
they derive a mean age of the low-mass stellar population of 4--5 Myr.

Modern CCD investigations of NGC\,2264 have yielded similar ages, but they are strongly dependent upon the pre-main
sequence models adopted for use. Sung et al. (1997) use the pre-main sequence models of Bernasconi \& Maeder (1996)
and Swenson et al. (1994) to find that the ages of most suspected pre-main sequence members of NGC\,2264 are from 0.8 to 8 Myr,
while the main sequence stars range from 1.4 to 16 Myr. Park et al. (2000) compare ages and age dispersions of NGC\,2264
from four sets of pre-main sequence models: those of Swenson et al. (1994), D'Antona \& Mazzitelli (1994), Baraffe et al. (1998),
and the revised models of Baraffe et al. (1998), which incorporate a different ratio of mixing length to pressure scale height.
From the distribution of pre-main sequence candidates, the median ages and age dispersions (respectively) from the models of
Swenson et al. (1994) are 2.1 and 8.0 Myr; D'Antona \& Mazzitelli (1994) 0.9 and 5.5 Myr; Baraffe et al. (1998) 4.3 and 15.3 Myr,
and for the revised models of Baraffe et al. (1998) 2.7 and 10 Myr. Figure 4 is a reproduction of the HR diagrams from Park et al. (2000),
with the evolutionary tracks for the various models superposed. The cluster ages and age spreads are represented by the solid and
dashed lines, respectively. These surveys used photometry alone to place stars on the HR diagram.

\begin{figure}
\centering
\includegraphics[angle=0,width=\linewidth, draft=False]{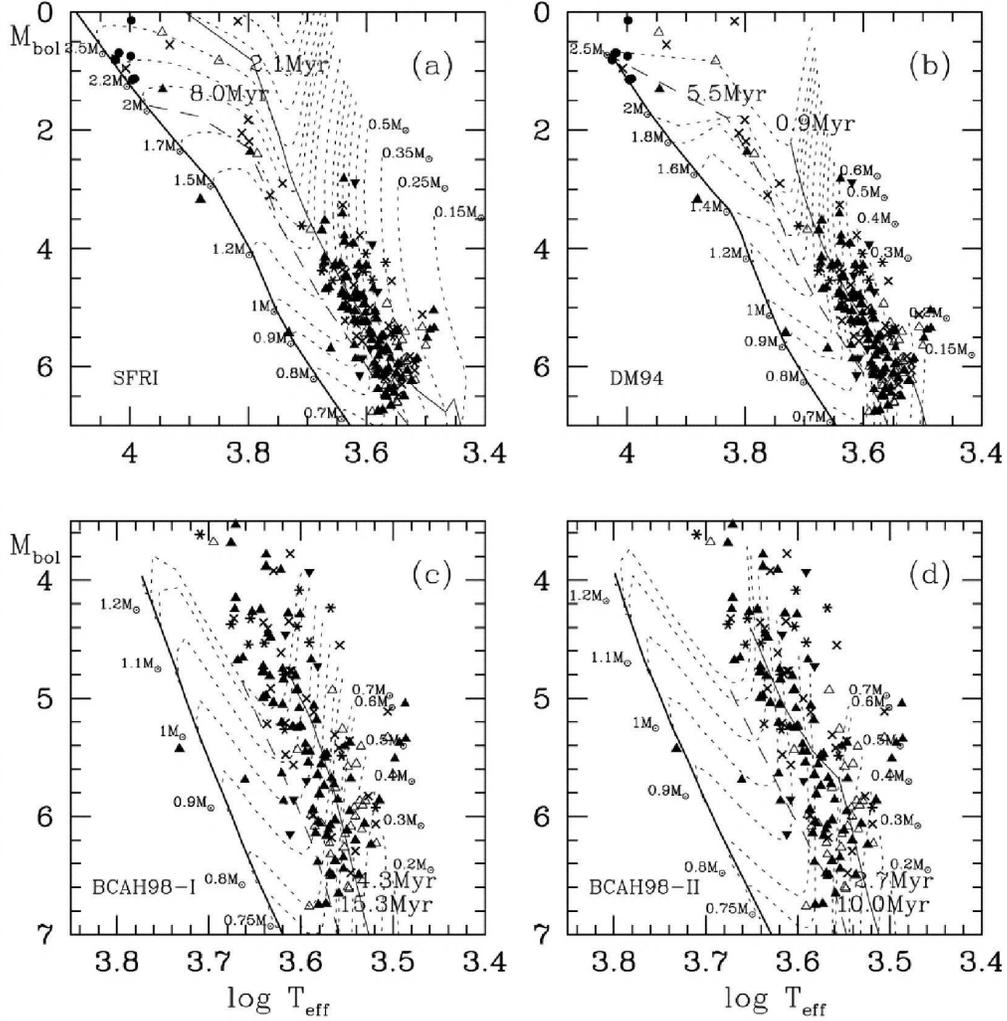}
\caption[fg4.eps]{The HR diagram of NGC\,2264 from Park et al. (2000) comparing the pre-main sequence models of
Swenson et al. (1994), D'Antona \& Mazzitelli (1994), Baraffe et al. (1998), and the revised models of Baraffe et al. (1998).
The thick solid lines represent the ZAMS, the dotted lines are the various evolutionary tracks, and the ages and age dispersions
are shown by the solid and dashed lines, respectively. \label{f4}}
\end{figure}

Rebull et al. (2002) compare the derived ages and masses from D'Antona \& Mazzitelli (1994) and
Siess et al. (2000) for a spectroscopically classified sample of stars in
NGC\,2264, finding systematic differences between the models of up to a factor of two in mass
and half an order of magnitude in age. Dahm \& Simon (2005) use the evolutionary models of D'Antona \& Mazzitelli (1997)
to determine a median age of 1.1 Myr and to infer an age dispersion of $\sim$5 Myr for the nearly
500 H$\alpha$ emission stars identified within the immediate cluster vicinity. Star formation, however,
continues within the cluster as evidenced by {\it Spitzer} observations of the star forming cores near IRS1 and IRS 2.
Young et al. (2006) suggest that a group of embedded, low-mass protostars coincident with IRS2 exhibits a velocity dispersion
consistent with a dynamical age of several $10^{4}$ yr. The presence of this young cluster as well
as other deeply embedded protostars (Allen 1972; Castelaz \& Grasdalen 1988; Margulis et al. 1989;
Thompson et al. 1998; Young et al. 2006) among the substantial number of late B-type dwarfs in NGC\,2264
implies that an intrinsic age dispersion of at least 3--5 Myr exists among the cluster population.
Soderblom et al. (1999) obtained high resolution spectra for 35 members of NGC\,2264 in order to determine
Li abundances, radial velocities, rotation rates, and chromospheric activity levels. Their radial velocities
indicate that the eight stars in their sample lying below the 5 Myr isochrone of the cluster are non-members,
implying that the age spread within the cluster is only $\sim$4 Myr. The hierarchical structure of the cluster
would indicate that star formation has occurred in different regions of the molecular cloud over the last
several Myr. We can speculate that from the large quantities of molecular gas remaining within the various
cloud cores, star formation will continue in the region for several additional Myr.
The cluster ages adopted or derived by selected previous investigations of NGC\,2264 are also presented in Table 1.

\section{The OB Stellar Population of NGC\,2264}

 S~Mon (15~Mon) dominates the northern half of the cluster and is believed to be the ionizing source for the Cone
Nebula (Schwartz et al. 1985) as well as many of the observed bright rims in the region including
Sharpless 273. In addition to exhibiting slight variability (hundredths of a magnitude), Gies et al.
(1993, 1997) determined S~Mon to be a visual and spectroscopic binary from speckle interferometry,
{\it HST} imaging, and radial velocity data. With a semi-major axis of over 27 AU (assuming a
distance of 800 pc), the 24-year orbit of the binary is illustrated in Figure 5, taken from
Gies et al. (1997) and based upon an orbital inclination of 35$^{\circ}$ to the plane of the sky. The mass estimates for the
primary component of S~Mon (O7 V), 35 M$_{\odot}$, and the secondary (O9.5 V), 24 M$_{\odot}$, assume a distance of 950 pc
(P\'{e}rez et al. 1987), which is significantly greater than the currently accepted cluster
distance. Using their derived orbital elements and with the adopted cluster distance (800 pc), a
mass ratio of q$=$0.75 is derived, leading to primary and secondary masses 18.1 and 13.5 M$_{\odot}$,
respectively. These mass estimates, however, are inconsistent with the spectral type of S~Mon and its
companion. Over the last 13 years, observations of the star using the fine guidance sensor onboard HST
have revised earlier estimates of the orbital elements, and the period of the binary now appears to be
significantly longer than previously determined (Gies, private communication).
S~Mon also exhibits UV resonance line profile variation as well as fluctuation in soft X-ray flux (Snow et al. 1981; Grady et al. 1984).
Both of these phenomenon are interpreted as being induced by variations in mass loss rate.

\begin{figure}
\centering
\includegraphics[angle=0,width=3.0in, draft=False]{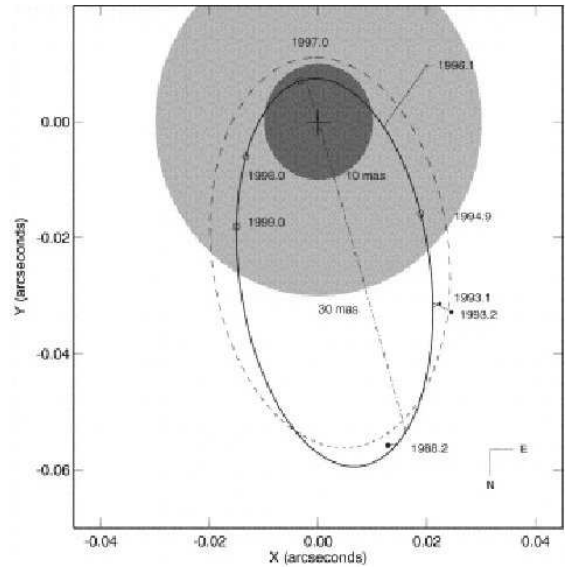}
\caption[fg5.eps]{The relative orbit of S~Mon B from Gies et al. (1997). The thick solid line represents the
orbital solution, the thin lines connect the observed and predicted positions.  The shaded circle represents
the resolution limit of a 4-meter class telescope. The dot-dashed line indicates the line of the nodes. \label{f5}}
\end{figure}

The OB-stellar population of NGC\,2264 consists of at least two dozen stars, several of which were spectroscopically
classified by Walker (1956). Morgan et al. (1965) list 17 early-type stars within the cluster, ranging in spectral type
from O7 to B9. Most of these stars are concentrated around S~Mon or the small rosette-shaped nebulosity lying to its
southwest, however, several do lie in the southern half of the cluster. Table 2 lists the suspected early-type members
of NGC\,2264 with their positions, available photometry, spectral types, and any relevant notes of interest. Several
members of the greater Mon OB 1 association are also included in Table 1 for completeness, some obtained from the
compilation of Turner (1976). The Hipparcos survey of de Zeeuw et al. (1999) was unable to provide any rigorous kinematic
member selection for Mon OB1 members given the distance of the association. Many of the early-type stars are known binaries
including S~Mon, HD 47732, and HD 47755 (Gies et al. 1993, 1997; Morgan et al. 1965; Dahm et al. 2007).
The B3 V star HD 47732 is identified by Morgan et al (1965) as a double-line
spectroscopic binary and is examined in detail by Beardsley \& Jacobsen (1978) and Koch et al. (1978). Another early-type
cluster member, HD 47755, was classified by Trumpler (1930) as a B5-type spectroscopic binary, but assigned a low probability of
membership by the proper motion survey of Vasilevskis et al. (1965). Koch et al. (1986) derive a 1.85 day period
for the shallow 0.1 mag eclipse depth. Dahm et al. (2007) list several early-type adaptive optics (AO) binaries and give
their separations and position angles (see their Table 2).
Other massive cluster members include several mid-to-late B-type stars clustered near
S~Mon including BD$+$10$^{\circ}$1222 (B5V), HD 261938 (B6V), W 181 (B9V), HD 261969 (B9IV), and $\sim$7\arcmin\ to the east,
HD 47961 (B2V). Although on the cloud periphery, HD 47961 is surrounded by several X-ray and H$\alpha$ emission stars and its
derived distance modulus, m$-$M$_{V}$$=$9.74, matches that of the cluster. The proper motion data of Vasilevskis et al. (1965)
also support its cluster membership status.

Within the rosette-shaped emission/reflection nebula southwest of S~Mon is a second clustering of massive stars including
W 67 (B2V), HD 47732 (B3V), HD 47755 (B5V), W 90 (B8e), and HD 261810 (B8V).  Between this grouping and HD 47887 (W 178),
the B2 III binary just north of  the Cone Nebula, there are just a handful of candidate early-type members: HD 47777 (B3V),
HD 262013 (B5V), and HD 261940 (B8V). HD 47887 was assigned a low probability of membership by Vasilevskis et al. (1965),
but if assumed to lie on the ZAMS, the distance modulus of this B2 binary equals that of the cluster. The other component
of HD 47887 is the B9 dwarf, HD 47887B, several arcseconds southwest of the primary. HD 262042, the B2 dwarf directly south of
the Cone Nebula, is a probable background star given its distance modulus ($m-M_{V} = 10.65$). Far to the west is the B5 star
HD 47469, a probable non-member lying $\sim$200 pc in front of the cluster. Most of the OB stellar population of the cluster
lies concentrated within the boundaries of the molecular cloud, matching the distribution of suspected low-mass members
(i.e. X-ray detected sources and H$\alpha$ emitters).

Walker (1956) notes the presence of five yellow giants within the cluster region and suggests they might lie in the foreground.
He assigns to these potential interlopers spectral types and luminosity classes of G5 III to W73, K2 II-III to W237, K3 II-III
to W229, K3 II-III to W69, and K5 III to {W37}. Underhill (1958) presents radial velocities of these stars, finding that only one
(W73) is consistent with the mean radial velocities of the early-type stars.

\begin{table}[tb]
\begin{small}
\begin{center}
\begin{tabular}[c]{ccccccc}
\multicolumn{7}{c}{Table 2} \\
\multicolumn{7}{c}{OB Stars in Proximity of NGC\,2264} \\
\tableline
Identifier     & RA (J2000)  & $\delta$ (J2000) & $V$   & $B-V$ & Sp T &  Notes\tablenotemark{a}\\
\tableline
HD 44498       & 06 22 22.5 & +08 19 36 &  8.82 & $-0.06$  & B2.5V  & Mon OB1\\
HD 45789A      & 06 29 55.9 & +07 06 43 &  7.10 & $-0.13$  & B2.5IV &        \\
HD 45827       & 06 30 05.5 & +09 01 46 &  6.57 & 0.11     & A0III  &        \\
HD 46300       & 06 32 54.2 & +07 19 58 &  4.50 & 0.00     & A0Ib   & 13 Mon \\
HD 46388       & 06 33 20.3 & +04 38 58 &  9.20 & 0.15     & B6V   &  Mon OB1\\
HD 261490      & 06 39 34.3 & +08 21 01 &  8.91 & 0.18     & B2III &  Mon OB1\\
HD 261657      & 06 40 04.8 & +09 34 46 & 10.88 & 0.03     & B9V  &          \\
HD 47732       & 06 40 28.5 & +09 49 04 &  8.10 & $-0.12$  & B3V  &  bin     \\
BD+09 1331B    & 06 40 37.2 & +09 47 30 & 10.79 & 0.62     & B2V  &          \\
HD 47755       & 06 40 38.3 & +09 47 15 &  8.43 & $-0.13$  & B5V  &  EB      \\
HD 47777       & 06 40 42.2 & +09 39 21 &  7.95 & $-0.17$  & B3V  &          \\
HD 261810      & 06 40 43.2 & +09 46 01 &  9.15 & $-0.13$  & B8V  &  var     \\
V V590 Mon     & 06 40 44.6 & +09 48 02 & 12.88 & 0.15     & B8pe  &  W90\\
BD+10 1220B    & 06 40 58.4  & +09 53 42   &  7.6  & ...      & OB    & S~Mon B\\
HD 261902      & 06 40 58.5 & +09 33 31 & 10.20 & 0.06     & B8V  &  \\
HD 47839       & 06 40 58.6 & +09 53 44 &  4.66 & $-0.25$  & O7Ve &  S~Mon A\\
BD+10 1222     & 06 41 00.2 & +09 52 15 &  9.88 & $-0.07$  & B5V  &  var \\
HD 261938      & 06 41 01.8 & +09 52 48 &  8.97 & $-0.07$  & B6V  &  var \\
HD 261903      & 06 41 02.8 & +09 27 23 &  9.16 & $-0.08$  & B9V  &  \\
HD 261940      & 06 41 04.1 & +09 33 01 & 10.0  & 0.04     & B8V  &  \\
HD 261937      & 06 41 04.5 & +09 54 43 & 10.36 & 0.35     & B8V  &  bin\\
HD 47887       & 06 41 09.6 & +09 27 57 &  7.17 & $-0.20$  & B2III &  bin\\
HD 47887B      & 06 41 09.8  & +09 27 44   &  9.6  & ...      & B9V  &  \\
IRS 1          & 06 41 10.1 & +09 29 33 & ...   & ...      & B2--B5&   \\
HD 261969      & 06 41 10.3 & +09 53 01 &  9.95 & 0.00     & B9IV &  var \\
W 181          & 06 41 11.2 & +09 52 55 & 10.03 & $-0.05$  & B9Vn &  bin\\
HD 262013      & 06 41 12.9 & +09 35 49 &  9.34 & $-0.12$  & B5Vn &  \\
HD 262042      & 06 41 18.7 & +09 12 49 &  9.02 & 0.03     & B2V  &  \\
HD 47934       & 06 41 22.0 & +09 43 51 &  8.88 & $-0.15$  & B9V  &  \\
HD 47961       & 06 41 27.3 & +09 51 14 &  7.51 & $-0.17$  & B2V  &  bin\\
HD 48055       &  06 41 49.7 & +09 30 29 &  9.00 & $-0.15$  & B9V  & var \\
\tableline
\noalign{\smallskip}
\multicolumn{7}{l}{\parbox{0.8\textwidth}{\footnotesize
    $^{a}$ var - variable; bin - binary; EB - eclipsing binary}}\\
\end{tabular}
\end{center}
\end{small}
\end{table}

\section{The Intermediate and Low Mass Stellar Population of NGC\,2264}

\subsection{H$\alpha$ Emission Surveys of NGC\,2264}

The full extent of the intermediate and low-mass ($< 3$ M$_{\odot}$)
stellar population of NGC 2264 and its associated molecular cloud complex
has not been assessed, but several hundred suspected members exhibiting H$\alpha$ emission, X-ray emission,
or photometric variability have been cataloged principally in the vicinity of the known molecular cloud cores.
The traditional means of identifying young, low-mass stars has been the detection of H$\alpha$ emission either
using wide-field, low-resolution, objective prism imaging or slitless grating spectroscopy.
In pre-main sequence stars, strong H$\alpha$ emission is generally attributed to accretion processes
as gas is channeled along magnetic field lines from the inner edge of the circumstellar disk to the
stellar photosphere. Weak H$\alpha$ emission is thought to arise from enhanced chromospheric activity,
similar to that observed in field dMe stars. Herbig (1954) conducted
the first census of NGC\,2264, discovering 84 H$\alpha$ emission stars concentrated in two groups in a
$25\arcmin \times 35\arcmin$ rectangular area centered roughly between S~Mon and the Cone Nebula. The larger
of the two groups lies north of HD 47887, the bright B2 star near the tip of the Cone Nebula. The second group is
concentrated southwest of S~Mon, near W67 and the prominent rosette of emission/reflection nebulosity. Marcy (1980)
re-examined NGC\,2264 using the identical instrumentation of Herbig (1954) and discovered 11 additional H$\alpha$
emitters, but was unable to detect emission from a dozen of the original LH$\alpha$ stars. The implication was
that H$\alpha$ emission varied temporally among these low-mass stars, and that a reservoir of undetected emitters
was present within the cluster and other young star forming regions.

Narrow-band filter photometric techniques have also been used to identify strong H$\alpha$ emitters in NGC\,2264.
The first such survey, by Adams et al. (1983), was photographic in nature and capable of detecting
strong emitters, i.e. classical T Tauri stars (CTTS). Their deep $UBVRIH\alpha$ photographic survey of the cluster found $\sim$300
low-luminosity (7.5$<M_{V}<$12.5) pre-main sequence candidates. Ogura (1984) conducted an objective prism
survey of the Mon OB1 and R1 associations, focusing upon regions away from the core of NGC\,2264 that had been
extensively covered by Herbig (1954) and Marcy (1980). This Kiso H$\alpha$ survey discovered 135 new emission-line
stars, whose distribution follows the contours of the dark molecular cloud complex. More than three dozen
of these H$\alpha$ emitters lie within or near NGC\,2245 to the west. In the molecular cloud region north of NGC\,2264,
another two dozen emitters were identified by Ogura (1984). Reipurth et al. (2004b), however, noticed a significant
discrepancy with Ogura's (1984) emitters, many of which were not detected in their objective prism survey. It
is possible that many of Ogura's (1984) sources are M-type stars which exhibit depressed photospheric continua
near H$\alpha$ caused by strong TiO band absorption. The resulting ``peak'' can be confused for H$\alpha$ emission.
Sung et al. (1997) conducted a CCD narrow-band
filter survey of NGC\,2264, selecting pre-main sequence stars based upon differences in $R-H\alpha$ color with
that of normal main sequence stars. Using this technique, Sung et al. (1997) identified 83 pre-main
sequence stars and 30 pre-main sequence candidates in the cluster. Of these 83 pre-main sequence stars, Dahm \& Simon (2005)
later established that 61 were CTTSs, 12 were weak-line T Tauri stars (WTTS) and several could not be positively identified.
While capable of detecting strong H$\alpha$ emission, the narrow-band filter photometric techniques are not able
to distinguish the vast majority of weak-line emitters in the cluster. Lamm et al. (2004) also obtained narrow-band
H$\alpha$ photometry for several hundred stars in their periodic variability survey of NGC\,2264.

More recently, Reipurth et al. (2004b) carried out a 3$^{\circ}$$\times$3$^{\circ}$
objective prism survey of the NGC\,2264 region using the ESO 100/152 cm Schmidt telescope at La Silla, which
yielded 357 H$\alpha$ emission stars, 244 of which were newly detected. Reipurth et al. (2004b) ranked emission strengths
on an ordinal scale from 1 to 5, with 1 indicating faint emission and strong continuum and 5 the inverse. Comparing
with the slitless grism survey of Dahm \& Simon (2005), the technique adopted by Reipurth et al. (2004b)
is capable of detecting weak emission to a limit of W(H$\alpha$)$\sim5$\AA. Shown in Figure 6
(from Reipurth et al. 2004b) is the distribution of their 357 detected H$\alpha$ emitters, superposed
upon a $^{13}$CO (1--0) map of the region obtained with the AT\&T Bell Laboratories 7 m offset
Cassegrain antenna. The main concentrations of H$\alpha$ emitters were located within or near
the strongest peaks in CO emission strength. A halo of H$\alpha$ emitters in the outlying regions
does not appear to be associated with the CO cloud cores and may represent an older population of
stars that have been scattered away from the central molecular cloud. Reipurth et al. (2004b)
speculate that this may be evidence for prolonged star formation activity within the molecular cloud
hosting NGC\,2264.
The slitless wide-field grism spectrograph (WFGS) survey of the cluster by Dahm \& Simon (2005) increased
the number of H$\alpha$ emitters in the central $25\arcmin \times 40\arcmin$ region between S~Mon and the Cone Nebula to nearly
500 stars. Capable of detecting weak emission with {\it W}(H$\alpha$) $\ge 2$\AA, the WFGS observations added more than
200 previously unknown H$\alpha$ emitters to the growing number of suspected cluster members. Combining the number of
known emitters outside of the survey of Dahm \& Simon (2005) identified by Ogura (1984) and Reipurth et al. (2004b),
we derive a lower limit for the low-mass (0.2--2 M$_{\odot}$) stellar population of the cluster region of
more than 600 stars. From a compilation of 616 candidate cluster members with $I < 20$, (from two deep {\it Chandra}
observations), Flaccomio (private communication) finds that 202 do not exhibit detectable H$\alpha$ emission.
Accounting for lower masses below the detection threshold of the various H$\alpha$ surveys, X-ray selected stars
that lack detectable H$\alpha$ emission, and embedded clusters
of young stars still emerging from their parent molecular cloud cores, the stellar population of the cluster could exceed
1000 members.

\begin{figure}
\centering
\includegraphics[angle=0,width=0.95\textwidth, draft=False]{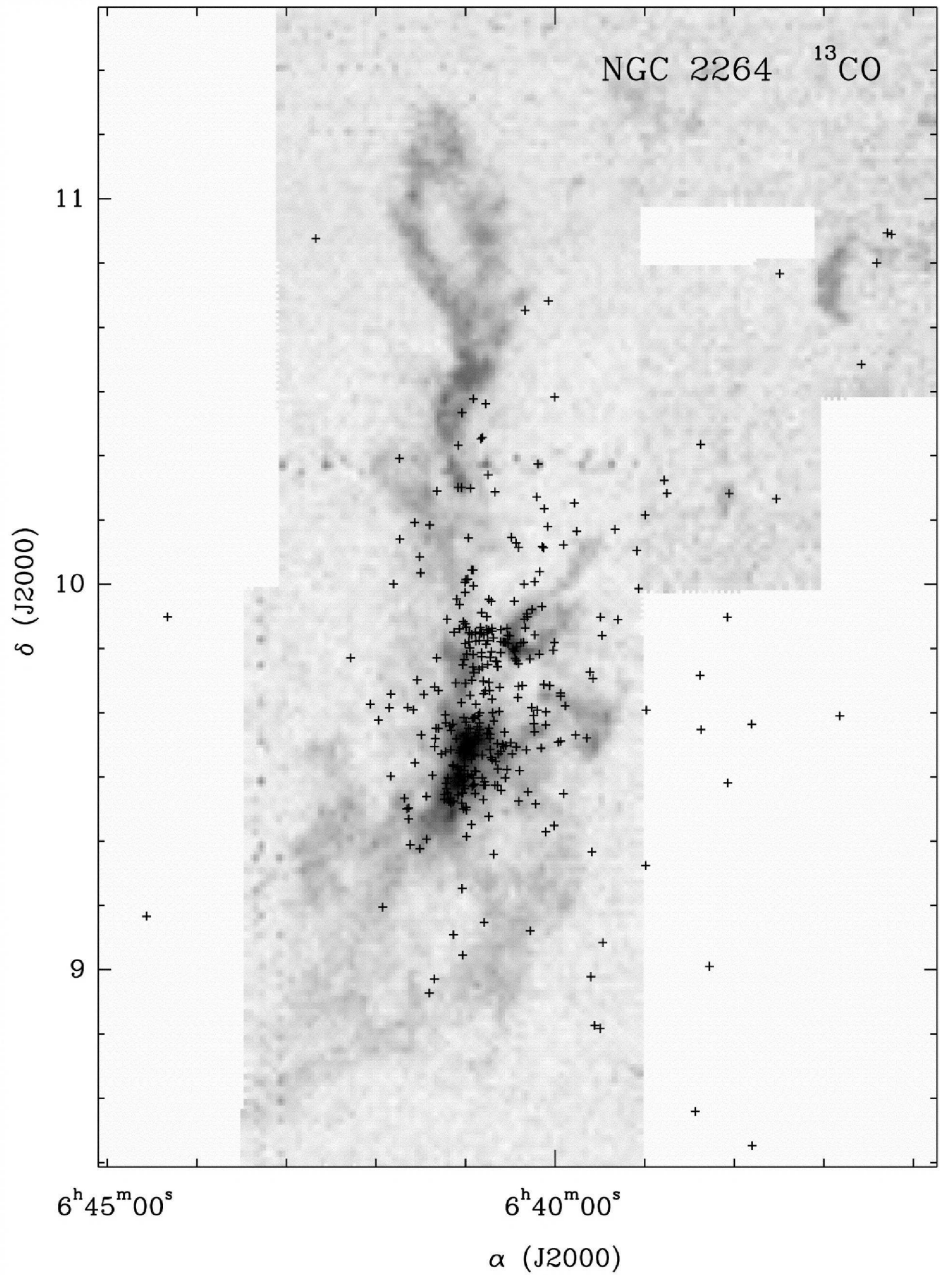}
\caption[fg6.eps]{The distribution of H$\alpha$ emitters in NGC\,2264 from a
3$^{\circ}$$\times$3$^{\circ}$ objective prism survey of the region by Reipurth et al. (2004b), superposed upon a grayscale map of
$^{13}$CO (1--0) by John Bally. The emitters are concentrated along the broad ridgeline of the molecular cloud and near
S~Mon. North and west of the cluster, H$\alpha$ emitters are scattered around and within the molecular cloud
perimeter. The stars that appear unassociated with molecular emission may have been scattered from the central
cluster region, implying that star formation has been occurring with the molecular cloud complex for several
Myr.
 \label{f6}}
\end{figure}

\subsection{X-ray Selected Members of NGC\,2264}
 X-ray emission is another well-established discriminant of stellar youth,
with pre-main sequence stars often exhibiting X-ray luminosities ($L_{\rm X}$) 1.5 dex
greater than their main sequence counterparts. The importance of X-ray emission as a
youth indicator was immediately recognized in young clusters and associations where field
star contamination is significant. The ability to select pre-main sequence stars from
field interlopers, most notably after strong H$\alpha$ emission has subsided, led to the
discovery of a large population of weak-line T Tauri stars associated with many nearby star forming
regions. Among field stars, however, the e-folding time for X-ray emission persists well into the main
sequence phase of evolution. Consequently, some detected X-ray sources are potentially foreground
field stars not associated with a given star forming region. For NGC\,2264, Flaccomio (private communication)
estimates the fraction of {\it Chandra} detected X-ray sources that are foreground interlopers to be $\sim$7\%.

Detailed discussion of X-ray observations made to date with ROSAT, {\it XMM-Newton},
or {\it Chandra} is given in Section 9, but here we summarize these results in the context of the low-mass
stellar population of NGC\,2264. Patten et al. (1994) used the High Resolution Imager (HRI) onboard ROSAT
to image NGC\,2264 and identified 74 X-ray sources in the cluster, ranging from S~Mon to late-K type pre-main
sequence stars. This shallow survey was later incorporated by Flaccomio et al. (2000) who combined six pointed
ROSAT HRI observations of the cluster for a total integration time of 25 ks in the southern half and 56 ks in
the northern half of the cluster. In their sample of 169 X-ray detections (133 with optical counterparts),
$\sim75$ were GKM spectral types, including many known CTTSs and WTTSs. Ramirez et al. (2004) cataloged
263 X-ray sources in a single 48.1 ks observation of the northern half of the cluster obtained with the Advanced
CCD Imaging Spectrometer (ACIS) onboard {\it Chandra}. Flaccomio et al. (2006) added considerably to the X-ray
emitting population of the cluster in their analysis of a single 97 ks {\it Chandra} ACIS-I observation of the
southern half of the cluster, finding an additional 420 sources. Of the nearly 700 X-ray sources detected by
{\it Chandra} in the cluster region,  509 are associated with optical
or NIR counterparts. Of these, the vast majority are still descending their convective tracks.
Dahm et al. (2007) identify 316 X-ray sources in NGC\,2264 in two
42 ks integrations of the northern and southern cluster regions made with the European Photon Imaging Camera
(EPIC) onboard {\it XMM-Newton}. The vast majority of these sources have masses $\le$ 2 M$_{\odot}$.
The positions of these X-ray selected stars as well as several embedded sources lacking optical
counterparts are shown in Figure 7, superposed upon the red DSS image of the cluster.

\begin{figure}[!tbh]
\centering
\includegraphics[angle=0,width=5in, draft=False]{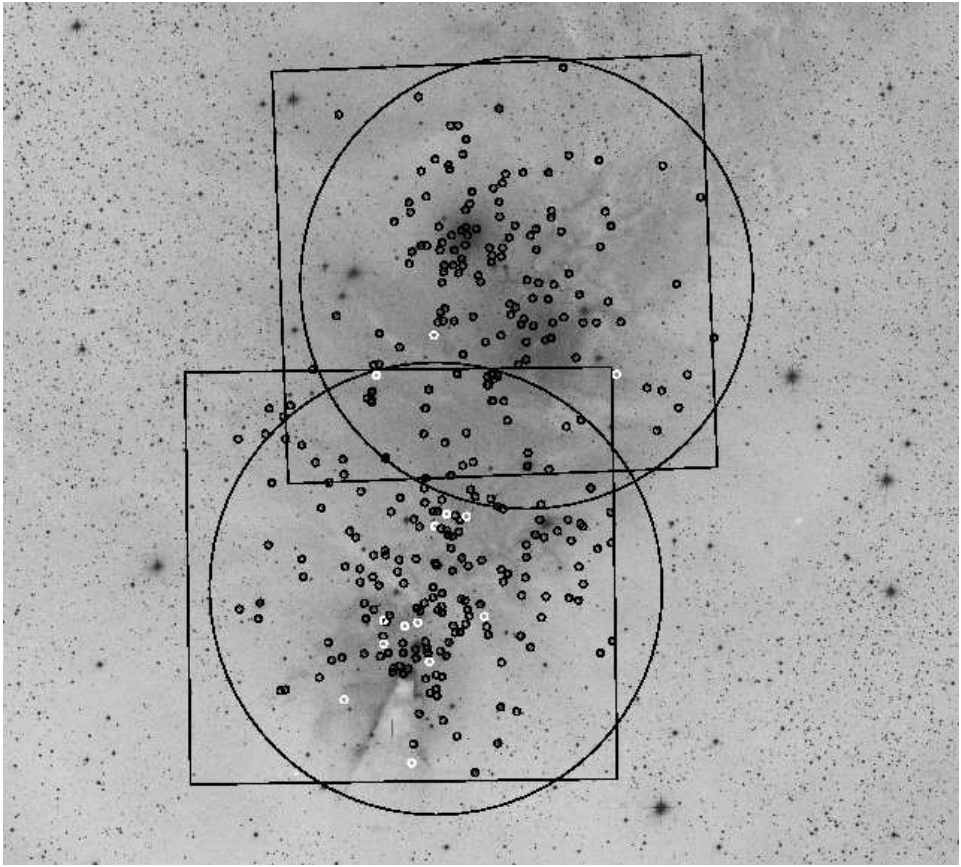}
\caption[fg7.eps]{The 1$^{\circ}$ square red image of NGC\,2264 obtained from the Digitized Sky Survey (DSS)
showing the field of view for each of the {\it XMM} EPIC exposures from the survey of Dahm et al. (2007).
The black circles mark 300 detected X-ray sources having optical counterparts, presumably cluster members,
while the white circles denote embedded sources or background objects observed through the molecular cloud.
The distribution of X-ray sources closely matches the distribution of H$\alpha$ emitters seen in Figure 6.
\label{f7}}
\end{figure}

\subsection{Photometric Variability Surveys}

 Photometric variability is among the list of criteria established by Joy (1945)
for membership in the class of T Tauri stars. Young, low-mass stars are believed to experience
enhanced, solar-like magnetic activity in which large spots or plage regions cause photometric
variability at amplitudes typically $<$0.2 mag. If active accretion is still taking place from
a circumstellar gas disk, hot spots from the impact point on the stellar surface are believed
to induce large amplitude, irregular variations.
Wolf (1924) first examined NGC\,2264 for photometric variables, identifying 24 from nine
archived plates of the cluster taken between 1903 and 1924. Herbig (1954) noted that 75\% (18 of 24)
of these variables also exhibited H$\alpha$ emission. Herbig (1954) further suggested that given
the variable nature of H$\alpha$ emission, continued observation of the cluster would likely remove
some of the variables from the non-emission group of stars. Nandy (1971) found that among
the H$\alpha$ emitters in NGC\,2264 that are also photometric variables, a positive correlation
exists between infrared (out to $I-$band) and ultraviolet excesses. Nandy \& Pratt (1972) show
that the range of variability among the T Tauri population in NGC\,2264 is less in $I-$band
than in other filters ($UBVR$) and also demonstrate a technique of studying the extinction
properties of dust grains enveloping T Tauri stars from their color variations.
Breger (1972) undertook an extensive photometric variability study of NGC\,2264,
finding that only 25\% of the pre-main sequence A and F-type stars exhibit short-period
variations. Breger (1972) also found a strong correlation between ``shell indicators''
and variability, particularly among the T Tauri stars. W90 (LH$\alpha$ 25) was also found
to have brightened by $\sim0.5$ mag in $V$ since 1953, an interesting trend for this
peculiar Herbig AeBe star. Koch \& Perry (1974) undertook an extensive photographic study of
the cluster, identifying over 50 new variables, 16 of which were suspected eclipsing binaries,
but no periods could be established. Koch \& Perry (1974) conclude that many more low-amplitude
($<$0.2 mag) variables must remain unidentified within the cluster, a prescient statement
that would remain unproven until the coming of modern CCD photometric surveys.
Variability studies of individual cluster members include those of Rucinski (1983) for W92,
Koch et al. (1978) for HD 47732, and  Walker (1977) also for W92.

CCD photometric monitoring programs have proven highly effective at identifying pre-main sequence
stars from variability analysis. Kearns et al. (1997) reported the discovery of nine periodic
variables in NGC\,2264, all G and K-type stars. An additional 22 periodic variables were identified
by Kearns \& Herbst (1998) including the deeply eclipsing (3.5 mag. in $I$), late K-type star, KH-15D.
Two large-scale photometric variability campaigns have added several hundred more periodic variables to those
initially identified by Kearns et al. (1997) and Kearns \& Herbst (1998). The surveys of Makidon et al. (2004) and Lamm et al. (2004)
were completed in the 2000-2001 and 1996-1997 observing seasons, respectively.  Their derived periods for
the commonly identified variables show extraordinary agreement, confirming the robustness of the technique.
The Makidon et al. (2004) survey used the 0.76 m telescope at McDonald Observatory imaging upon a single
2048$\times$2048 CCD.  The short focal length of the set-up resulted in a total field of view of approximately
46\arcmin square.  Their observations were made over a 102-day baseline with an average of 5 images per
night on 23 separate nights.  Makidon et al. (2004) identified 201 periodic variables in a period distribution
that was indistinguishable from that of the Orion Nebula Cluster (Herbst et al. 2002). This surprising result suggests that
spin-up does not occur over the age range from $<$1 to 5 Myr and radii of 1.2 to 4 R$_{\odot}$ (Rebull et al. 2004).
Makidon et al. (2004) also found no conclusive evidence for correlations between rotation period and the presence
of disk indicators, specifically $U-V$, $I_{c}-K$, and $H-K$ colors as well as strong H$\alpha$ emission.

The Lamm et al. (2004) survey was made using the Wide-Field Imager (WFI) on the MPG/ESO 2.2 m telescope at La Silla.
The WFI is composed of eight 2048$\times$4096 CCDs mosaiced together, resulting in a field of view of $\sim$33\arcmin\ square.
This $VR_{c}I_{c}$ survey was undertaken over 44 nights with between one and 18 images obtained per night.
The resulting photometric database from the Lamm et al. (2004) survey is one of the most extensive available for NGC\,2264
and includes spectral types from Rebull et al. (2002) as well as other sources.  Over 10,600 stars were monitored
by the program, of which 543 were found to be periodic variables and 484 irregular variables with 11.4$<$$I_{c}$$<$19.7.
Of these variables, 405 periodic and 184 irregulars possessed other criteria that are indicative of pre-main sequence stars.
Lamm et al. (2004) estimate that 70\% of all pre-main sequence stars in the cluster ($I_{c}$$<$18.0, $R_{c}-I_{c}$$<$1.8)
were identified by their monitoring program (implying a cluster population of $\sim$850 stars). Lamm et al. (2005) use
this extensive study to conclude that the period distribution in NGC\,2264 is similar to that of the Orion Nebula Cluster,
but shifted toward shorter periods (recall that no difference was noted by Makidon et al. 2004).
For stellar masses less than 0.25 M$_{\odot}$, the distribution is unimodal, but for more massive stars,
the period distribution is bimodal. Lamm et al. (2005) suggest that a large fraction of stars in NGC\,2264 are
spun up relative to their counterparts in the younger Orion Nebula Cluster. No significant spin up, however, was
noted between older and younger stars within NGC\,2264. Lamm et al. (2005) also find evidence for disk locking
among 30\% of the higher mass stars, while among lower mass stars, disk locking is less significant. The locking
period for the more massive stars is found to be $\sim$8 days. For less massive stars a peak in the period
distribution at 2--3 days suggests that these stars while not locked have undergone moderate angular momentum loss
from star-disk interaction.
Other photometric surveys of the low-mass stellar population of NGC\,2264 include those of P\'{e}rez et al. (1987, 1989)
and Fallscheer \& Herbst (2006), who examined $UVI$ photometry of 0.4--1.2 M$_{\odot}$ members and found a significant
correlation between $U-V$ excess and rotation such that slower rotators are more likely to exhibit ultraviolet excess.

\subsection{Spectroscopic Surveys of NGC\,2264}
 Numerous spectroscopic surveys of the intermediate and low-mass stellar populations of NGC\,2264 have been completed
including those of Walker (1954, 1956, 1972), Young (1978), Barry et al. (1979),
Simon et al. (1985), King (1998), Soderblom et al. (1999), King et al. (2000), Rebull et al. (2002), Makidon et al. (2004),
Dahm \& Simon (2005), and Furesz et al. (2006). Walker (1972) examined the spectra of 25 UV excess stars in the Orion Nebula Cluster and in
NGC\,2264 having masses between 0.2--0.5 M$_{\odot}$ and bolometric magnitudes ranging from $+2.5$ to $+5.8$.
From the radial velocities of the H I and Ca II lines, Walker (1972) concluded that nine of these stars were
experiencing accretion either from an enveloping shell of gas or from a surrounding disk. The rate of infall
was estimated to be $\leq$ 10$^{-6}$ M$_{\odot}$ yr$^{-1}$. Young (1978) presented spectral types for 69 suspected
cluster members ranging from B3V to K5V. Combining the spectra with broad-band photometry, he concluded that
differential reddening is present within the cluster, likely induced by intracluster dust clouds.
King (1998) examined Li features among six F and G-type stars within the cluster, finding the mean
non-LTE abundance to be log N(Li)$=3.27 \pm 0.05$. This is identical to meteoritic values within our
solar system, implying that no Galactic Li enrichment has occurred over the past 4.6 Gyr. From
rotation rates of 35 candidate members of NGC\,2264, Soderblom et al. (1999) conclude that the
stars in NGC\,2264 are spun-up relative to members of the Orion Nebula Cluster. Furesz et al. (2006) present
spectra for nearly 1000 stars in the NGC\,2264 region from the Hectochelle multiobject spectrograph on the MMT.
Of these, 471 are confirmed as cluster members based upon radial velocities or H$\alpha$ emission. Furesz et al. (2006)
also find that spatially coherent structure exists in the radial velocity distribution of suspected members,
similar to that observed in the molecular gas associated with the cluster.

\section{The Substellar Population of NGC\,2264}

 The substellar population of NGC\,2264 remains relatively unexplored given the distance of the cluster.
Several deep imaging surveys have been conducted of the region, including that of Sung et al. (2004)
and Kendall et al. (2005). Sung et al. (2004) observed NGC\,2264 with the CFHT 12K array in the
broadband $VRI$ filters and a narrowband H$\alpha$ filter. Their deep survey extends to the substellar limit.
The optical photometry was combined with {\it Chandra} X-ray observations of the cluster to derive the
cluster IMF. Using the ($R-I, V-I$) color-color diagram to remove background giants, Sung et al. (2004)
conclude that the overall shape of the cluster IMF is similar to that of the Pleiades or the Orion
Nebula Cluster.

Kendall et al. (2005) identified 236 low-mass candidates lying redward of the 2 Myr isochrones of the
DUSTY evolutionary models generated by Baraffe et al. (1998). Of these substellar candidates,
79 could be cross-correlated with the 2MASS database, thereby permitting dereddening from NIR excesses.
Most of these candidates range in mass between that of the Sun and the substellar limit.
The reddest objects with the lowest A$_{V}$ values are possible brown dwarfs, but
deeper optical and NIR ($K\sim18.5$) surveys are needed to further probe the substellar
population of NGC\,2264. For the brown dwarf candidates identified thus far within the
cluster, spectroscopic follow-up is needed for confirmation. Shown in Figure 8 is the ($I, I-Z$)
color-magnitude diagram from Kendall et al. (2005), which shows all 236 low-mass candidates identified
in the cluster.

\begin{figure}[tb]
\centering
\includegraphics[angle=0,width=0.9\textwidth, draft=False]{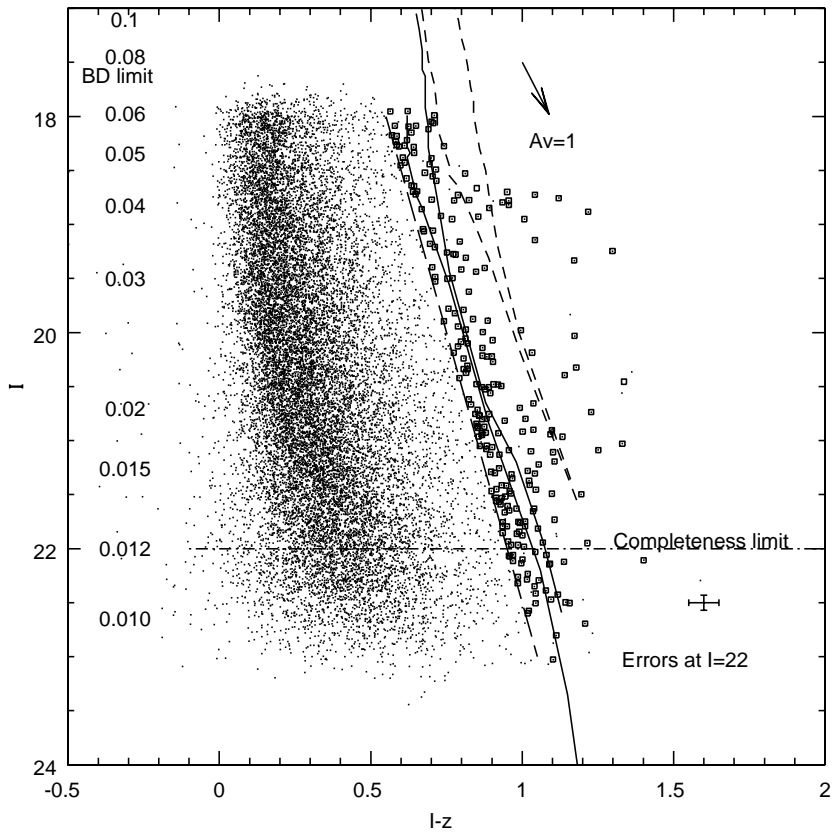}
\caption[fg8.SR.eps]{The ($I-Z$, $I$) color-magnitude diagram of NGC\,2264 from the survey of Kendall et al. (2005). Superposed
are the solid isochrones of the DUSTY models of Baraffe et al. for ages of 2 and 5 Myr. The dashed isochrones are
the NextGen stellar models for the same ages. The 236 substellar candidates (squares) were selected to be redward
of the sloping straight line. Distance assumed for the isochrones is 760 pc. The mass scale on the left is for the
2 Myr models. \label{f8}}
\end{figure}

\section{Centimeter, Millimeter, and Submillimeter Surveys of NGC\,2264}

 Among the earliest radio surveys of NGC\,2264 is that of Menon (1956) who observed the region
at 21 cm and found a decrease in neutral hydrogen intensity toward the molecular cloud, leading
to speculation that the formation of molecular hydrogen depleted the region of neutral atoms.
Minn \& Greenberg (1975) surveyed the dark clouds associated with the cluster at both the 6 cm
line of H$_{2}$CO and the 21 cm line of H I. Minn \& Greenberg (1975) discovered that the H$_{2}$CO
line intensity decreased dramatically outside of the boundaries of the dark cloud, implying that
the molecule was confined within the visual boundaries of the cloud. Depressions within the 21 cm H I
profiles observed along the line of sight were also found to be well-correlated with the H$_{2}$CO line
velocities, but no further quantitative estimates were made.

Molecular gas dominates the Mon OB1 association, with stars accounting for less than a few percent of the
total mass of the cluster-cloud complex. Zuckerman et al. (1972) searched for HCN and CS molecular line
emission in the Cone Nebula, but later discovered their pointing was actually 4\arcmin\ north of the Cone.
Emission, however, was serendipitously discovered from the background molecular cloud, presumably the region
associated with IRS1. Riegel \& Crutcher (1972) detected OH emission from several pointings within NGC\,2264
including a position near that observed by Zuckerman et al. (1972). From the relative agreement among line
radial velocities, Riegel \& Crutcher (1972) concluded that the OH emission arises from the same general region
as that of the other molecular species. Mayer et al. (1973) also detected NH$_{3}$ emission at the position
reported by Zuckerman et al. (1972), finding comparable velocity widths among the NH$_{3}$, HCN and CS lines.
The measured velocity widths, however, were larger than the thermal Doppler broadening inferred from the
NH$_{3}$ kinetic temperatures. From this, Mayer et al. (1973) suggested that a systematic radial velocity
gradient exists over the region where IRS1 is located. Rickard et al. (1977) mapped 6 cm continuum
emission and H$_{2}$CO absorption over an extended region of NGC\,2264, finding that H$_{2}$CO absorption
toward the cluster was complex and possibly arises from multiple cloud components.

Blitz (1979) completed a CO (1--0) 2.6 mm survey of NGC\,2264 at a resolution of 8\arcmin\
and identified two large molecular clouds in the region with a narrow bridge of gas between them. One of the
clouds is centered upon NGC\,2264 and the other lies 2$^{\circ}$ northwest of the cluster and
contains several reflection nebulae (NGC\,2245 and NGC\,2247). Crutcher et al. (1978)
mapped the region in the J$=$1--0 lines of $^{12}$CO and $^{13}$CO at somewhat better resolution
(half-power beam diameter of $\sim 2\farcm$6). Shown in Figure 9 are their resulting maps for
$^{12}$CO and $^{13}$CO. The primary cloud core lying approximately 8\arcmin\ north of the Cone Nebula
is elongated with a position angle roughly parallel to the Galactic plane.
The peak antenna temperature of 22 K occurs over an extended region, which includes IRS1 (Allen's source) but is
not centered upon this luminous infrared source. Two less prominent peaks were identified by
Crutcher et al. (1978) about 4\arcmin\ south and 8\arcmin\ west of S~Mon.

\begin{figure}
\centering
\includegraphics[angle=0,width=3in, draft=False]{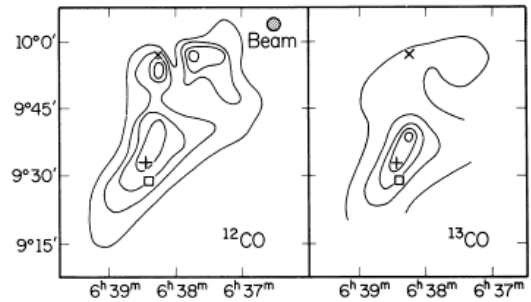}
\caption[fg9.ps]{Contour map for $^{12}$CO and $^{13}$CO of the NGC\,2264 region from Crutcher et al. (1978).
The coordinate scale is for equinox and epoch B1950.
The contour intervals are 4 and 2 K for the $^{12}$CO and $^{13}$CO maps, respectively, with the lowest contours
representing the 8 K isotherm. The cross, plus symbol, and square denote the positions of S~Mon, IRS1, and the Cone Nebula.
\label{f9}}
\end{figure}

The kinematic structure of the molecular clouds is somewhat
complex with a 2 km s$^{-1}$ velocity gradient across the primary cloud and three distinct velocity
components near S~Mon.  Individual clouds identified by Crutcher et al. (1978) are listed in Table 3,
reproduced from their Table 1. Included with their identifiers are the positions (B1950), radii in pc,
LSR velocity for $^{12}$CO, antenna temperatures for $^{12}$CO, and mass. Summing these individual
fragments, a total cloud mass of $\sim$7$\times$10$^{3}$ M$_{\odot}$ is estimated. This, however,
is a lower limit based upon CO column densities. Crutcher et al. (1978) also determined a virial
mass of 3$\times$10$^{4}$ M$_{\odot}$ for the molecular cloud complex from the velocities of the
individual cloud fragments. Assuming this to be an upper limit, Crutcher et al. (1978) adopt a
middle value of 2$\times$10$^{4}$ M$_{\odot}$ for the total cloud mass associated with the cluster,
half of which is contained within the primary cloud core north of the Cone Nebula. Considering the
energetics of the cloud, they further suggest that the luminous stars in NGC\,2264 are not
capable of heating the core of this massive cloud. This conclusion is challenged by Sargent et al. (1984) on
the basis of a revised cooling rate. Shown in Figure 10 is an optical image of the Cloud Core C region
from Crutcher et al. (1978), which lies just west of S~Mon.

\begin{figure}[tb]
\centering
\includegraphics[angle=0,width=0.9\textwidth, draft=False]{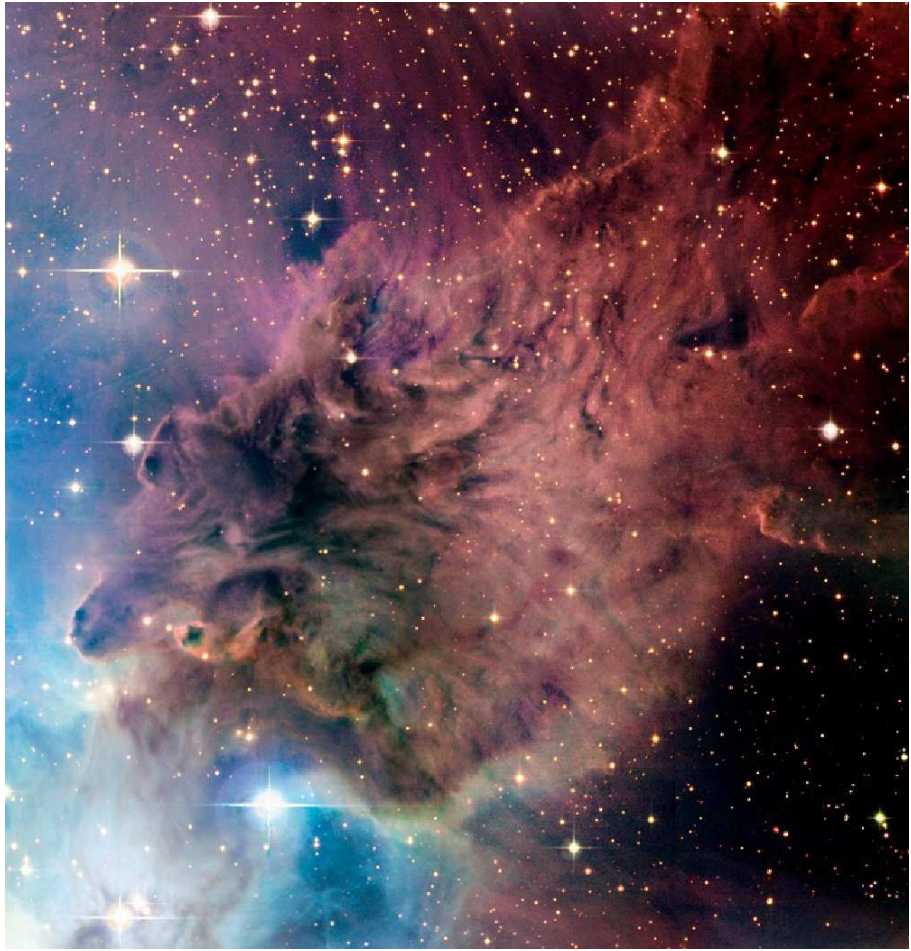}
\caption[fg10.eps]{Cloud Core C from Crutcher et al. (1978) is a cometary cloud located
immediately west of S Mon, and showing obvious effects of this proximity. The cloud core
is located in what is known in popular literature as the Fox Fur Nebula. The field is
approximately 15 $\times$ 15\arcmin, with north up and east left. Image obtained at the
CFHT. Courtesy J.-C. Cuillandre and G. Anselmi.
\label{f10}}
\end{figure}

\begin{table}
\begin{footnotesize}
\begin{center}
\begin{tabular}[c]{cccccccc}
\multicolumn{8}{c}{Table 3} \\
\multicolumn{8}{c}{Molecular Cloud Cores Associated with NGC\,2264 from Crutcher et al. (1978)} \\
\tableline
Identifier & $\alpha$              & $\delta$       &   R  &   V           & T$_{A}$  & N(CO)             & Mass \\
           & (B1950)         & (B1950)        & (pc) & (km/s) &  (K)     & 10$^{17}$ cm$^{-2}$  & (M$_{\odot}$)   \\
\tableline
A          & 6 37.6     & +9 37.0 & 1.7  & 4.9           &  12      &   8               & 1200   \\
B          & 6 37.2     & +9 56.0 & 0.9  & 8.8           &  19      &   8               & 400    \\
C          & 6 37.7     & +9 56.0 & 0.9  & 10.4          &  22      &   6               & 300    \\
D          & 6 37.9     & +9 49.0 & 0.4  & 11.4          &  14      &   2               & 20     \\
E          & 6 38.2     & +9 53.0 & 0.8  & 8.9           &  23      &  10               & 400    \\
F          & 6 38.2     & +9 40.0 & 0.8  & 5.5           &  22      &  53               & 1800   \\
G          & 6 38.4     & +9 32.0 & 0.9  & 7.5           &  22      &  54               & 2300   \\
H          & 6 38.9     & +9 22.0 & 0.9  & 4.5           &  11      &  5                & 300    \\
\tableline
\end{tabular}
\end{center}
\end{footnotesize}
\end{table}

Schwartz et al. (1985) completed CO (1--0), C$^{18}$O, and CS (3--2) CS observations of the IRS1
and IRS2 regions with the 11 meter NRAO antenna on Kitt Peak as well as NH$_{3}$ observations using
the NRAO 43 meter antenna at Green Bank. The NH$_{3}$ and CS observations were used to examine
high density gas while CO and C$^{18}$O were employed as tracers of H$_{2}$ column density.
Schwartz et al. (1985) combined these molecular gas observations with far-infrared (35--250~$\mu$m)
data obtained with the multibeam photometer system onboard the Kuiper Airborne Observatory.
The general morphology of the CO and CS maps of the regions were found to be similar, but two
key differences were noted: first, the IRS2 core appears as two unresolved knots in the CS map
and second, the high density gas near IRS1 (Allen's source) lies east of the far-infrared peak and appears extended
along the northeast to southwest axis. Schwartz et al. (1985) conclude that Allen's source is an
early-type star ($\sim$ B3V), which is embedded within a dense molecular cloud. They further
speculate that the molecular cloud associated with IRS1 is ring-shaped, lying nearly edge-on
with respect to the observer (see their Figure 7). This conclusion was drawn from the gas-density
and velocity structure maps of the region.

Krugel et al. (1987) completed a more spatially resolved survey of the IRS1 region
in both NH$_{3}$ and CO, mapping a 10\arcmin$\times$10\arcmin\ area. Two distinct cloud cores
were identified, each of roughly 500 M$_{\odot}$. Kinetic gas temperatures were found by the
authors to be 18 K in the north and 25 K in the south with a peak of 60 K around IRS1. Complex
structure was noted in the southern cloud, which exhibits multiple subclouds with differing
temperatures and velocities. Krugel et al. (1987) find that the smooth velocity gradient across
the region observed at lower resolution disappears completely in their higher resolution maps.
They also failed to detect high velocity wings toward IRS1 that might be associated with a
molecular outflow. Tauber et al. (1993) observed one of the bright rims in NGC\,2264
west of S~Mon in the $^{12}$CO and $^{13}$CO (3--2) transitions with the Caltech Submillimeter
Observatory. In their high spatial resolution maps (20\arcsec), they find a morphology which
suggests interaction with the ionizing radiation from S~Mon. The $^{13}$CO maps reveal a hollow
shell of gas broken into three components: two eastern clumps with their long axes pointing toward
S~Mon and a southern clump, which exhibits kinematic structure indicative of an embedded, rotating
torus of dense gas.

Oliver et al. (1996) completed a sensitive, unbiased CO (1--0) line emission survey of the Mon OB1 region
using the 1.2 meter millimeter-wave radio telescope at the Harvard Smithsonian Center for Astrophysics.
The survey consisted of over 13,400 individual spectra and extended from $l=196^{\circ}-206.5^{\circ}$,
with individual pointings uniformly separated by 3.75\arcmin. Oliver et al. (1996) find that the
molecular gas along the line of sight of the Mon OB1 association possesses radial velocities consistent with
the local spiral arm and the outer Perseus arm. Within the local arm, they identify 20 individual
molecular clouds ranging in mass from $\sim$100 to 5.2$\times$10$^{4}$ M$_{\odot}$. Their Table 5 (reproduced
here as Table 4) summarizes the properties of these molecular clouds including position, V$_{LSR}$, distance, radius,
mass, and associated clusters, molecular clouds, or bright nebulae. The most massive of these molecular
clouds hosts NGC\,2264, but their derived mass estimates assume the distance of P\'{e}rez et al. (1987),
950 pc, which probably overestimates the actual cluster distance by a factor of 1.2.
The revised cloud mass assuming a distance of 800 pc is 3.7 $\times$ 10$^{4}$ M$_{\odot}$,
 which is in somewhat better agreement with the value derived by Crutcher et al. (1978).
Oliver et al. (1996) also identify six arc-like molecular structures in the
Mon OB1 region, which may be associated with supernova remnants or wind-blown shells of gas. These
structures, if associated with the local spiral arm and at an appropriate distance, may imply that star
formation within the region was triggered by an energetic supernova event.\nopagebreak[4]

\begin{table}[tb]
\begin{footnotesize}
\begin{center}
\begin{tabular}[c]{c@{\hskip8pt}c@{\hskip8pt}c@{\hskip8pt}c@{\hskip8pt}c@{\hskip8pt}c@{\hskip8pt}c@{\hskip8pt}c@{\hskip4pt}c}
\multicolumn{9}{c}{Table 4} \\
\multicolumn{9}{c}{Properties of Observed Local Arm Molecular Clouds (Oliver et al. 1996)} \\
\tableline
Cloud & {\it l}              & {\it b}          &   V$_{lsr}$   &   $\delta$V    &   d   &  R     & Mass (CO) & Associations \\
           &  $^{\circ}$      & $^{\circ}$ &   km/s &   km/s  &  kpc  &  kpc   & M$_{\odot}$ &            \\
\tableline
1          &  196.25          &  $-$0.13   &   +4.7        &     4.6        &  0.9  &  9.4   & 3.7$\times$10$^{3}$ &     \\
2          &  196.75          &  +1.50     &   +5.3        &     2.3        &  0.9  &  9.7   & 9$\times$10$^{3}$   &     \\
3          &  196.88          &  $-$1.13   &   +5.3        &     3.3        &  0.9  &  9.4   & 1.1$\times$10$^{3}$ &     \\
4          &  198.88          &  +0.00     &   $-$6.9        &     3.4        &  0.8  &  9.3   & 1.0$\times$10$^{3}$ &     \\
5          &  199.31          &  $-$0.50   &   +3.3        &     2.6        &  0.9  &  9.4   & 6.7$\times$10$^{2}$ &     \\
6          &  199.56          &  $-$0.44   &   +6.4        &     4.9        &  0.9  &  9.4   & 2.0$\times$10$^{3}$ &     \\
7          &  199.81          &  +0.94     &   +6.9        &     4.6        &  0.9  &  9.4   & 2.4$\times$10$^{4}$ & L1600, L1601\\
           &                  &            &               &                &       &        &                     & L1604, LbN886\\
           &                  &            &               &                &       &        &                     & LbN889\\
8          &  200.19          &  +3.44     &   $-$6.8        &     3.4        &  0.8  &  9.3   & 1.9$\times$10$^{2}$ &       \\
9          &  200.81          &  +0.13     &  $-$11.0        &     5.2        &  0.8  &  9.3   & 1.5$\times$10$^{3}$ &  VY Mon, LbN895 \\
10         &  201.38          &  +0.31     &  $-$1.0        &     3.2        &  0.9  &  9.3   & 4.3$\times$10$^{3}$ &  Mon R1, L1605\\
           &                  &            &               &                &       &        &                     &  VDB76,77,78\\
           &                  &            &               &                &       &        &                     &  LbN898,903\\
11         &  201.44          &  +0.69     &   +5.1        &     4.4        &  0.9  &  9.3   & 2.1$\times$10$^{4}$ &  Mon R1, VY Mon\\
           &                  &            &               &                &       &        &                     &  NGC\,2245, NGC\,2247\\
           &                  &            &               &                &       &        &                     &  L1605, LbN901/904\\
12         &  201.44          &  +2.56     &   +0.7        &     3.1        &  0.9  &  9.4   & 2.0$\times$10$^{3}$ &  NGC\,2259\\
13         &  201.50          &  +2.38     &   +4.6        &     6.9        &  0.95 &  9.4   & 6.3$\times$10$^{3}$ &  NGC\,2259, LbN899\\
14         &  202.06          &  +1.44     &   +1.1        &     2.9        &  0.95 &  9.4   & 1.5$\times$10$^{3}$ &  L1609\\
15         &  202.25          &  +1.69     &   +4.7        &     2.5        &  0.95 &  9.4   & 9$\times$10$^{2}$   &  L1610\\
16         &  203.25          &  +2.06     &   +6.9        &     5.0        &  0.95 &  9.4   & 5.2$\times$10$^{4}$ &  NGC\,2264, IRS1, L1613\\
           &                  &            &               &                &       &        &                     &  S~Mon, LbN911/912/922\\
17         &  203.75          &  +1.25     &   +8.7        &     2.8        &  0.80 &  9.2   & 1.0$\times$10$^{3}$ &  NGC\,2261\\
           &                  &            &               &                &       &        &                     &  HH 39(A--F)\\
           &                  &            &               &                &       &        &                     &  R Mon, LbN920\\
18         &  204.13          &  +0.44     &   +5.2        &     2.8        &  0.80 &  9.2   & 1.9$\times$10$^{3}$ &  NGC\,2254, LbN929\\
19         &  204.44          &  $-$0.13   &  $-$3.1        &     2.6        &  0.80 &  9.2   & 1.1$\times$10$^{2}$ &  NGC\,2254\\
20         &  204.81          &  +0.44     &   +9.4        &     2.6        &  0.80 &  9.2   & 8.2$\times$10$^{2}$ &  NGC\,2254\\
\tableline
\end{tabular}
\end{center}
\end{footnotesize}
\end{table}

Wolf-Chase \& Gregersen (1997) analyzed observations of numerous transitions of CS and CO in the IRS1 region.
Taken as a whole, the observations suggest that gravitational infall is taking place.
Schreyer et al. (1997) mapped the region around IRS1 in various molecular transitions of CS, CO,
methanol, and C$^{18}$O.  Complementing the millimeter survey, Schreyer et al. (1997) also imaged
the region in the NIR ($JHK$), revealing the presence of several suspected low-mass stars surrounding
IRS1. To the southeast, a small, deeply embedded cluster was noted in the $K-$band mosaic of the
field, which coincided with a second cloud core identified by the millimeter survey. Extending
northwest from IRS1, Schreyer et al. (1997) noted a jet-like feature in all NIR passbands that
connects to the fan-shaped nebula evident on optical images of the field. Two bipolar outflows were
also detected in their CS mapping of the IRS1 region, one originating from IRS1 itself and another
from the millimeter source lying to the southeast. Thompson et al. (1998) examined IRS1 with NICMOS
onboard HST and discovered six point sources at projected separations of 2\farcs6 to 4\farcs9 from
Allen's source (IRS1). From the NIR colors and magnitudes of these sources, Thompson et al. (1998) suggest
that these faint stars are pre-main sequence stars whose formation was possibly triggered by the
collapse of IRS1.

Ward-Thompson et al. (2000) observed the IRS1 region in the millimeter and submillimeter continuum
(1.3 mm -- 350~$\mu$m), using the 7-channel bolometer array on the IRAM 30 meter telescope and the
UKT14 detector on the James Clerk Maxwell Telescope (JCMT) on Mauna Kea. The resulting maps revealed
a ridge of bright millimeter emission as well as a clustering of five sources with masses ranging
from 10--50 M$_{\odot}$. Table 4 of Ward-Thompson et al. (2000) summarizes the properties of these five
massive condensations, which are assumed to be the progenitor cores of intermediate-mass stars. The third
millimeter source (MMS3) of Ward-Thompson et al. (2000) was identified as the source of one of the bipolar
outflows discovered by Schreyer et al. (1997) in their millimeter survey of the region.

Williams \& Garland (2002) completed 870~$\mu$m continuum emission maps and (3--2) line surveys
of HCO$^{+}$ and H$^{13}$CO$^{+}$ of the IRS1 and IRS2 regions of NGC\,2264. The submillimeter
continuum emission was used to trace dust around the young clusters of protostars, while line emission
was used as a diagnostic of gas flow within the region. Shown in Figure 11 is the 870~$\mu$m
continuum emission map of Williams \& Garland (2002), their Figure 1, which clearly demonstrates
the fragmented nature of IRS2. Although IRS1 possesses a much higher peak flux density, the
masses of both regions are found to be comparable, $\sim$10$^{3}$ M$_{\odot}$. Williams \& Garland (2002)
find evidence for large-scale collapse for both IRS1 and IRS2 with infall velocities of $\sim$0.3 km s$^{-1}$.
From their derived virial mass of the IRS2 protostellar cluster, they conclude that the system is very
likely gravitationally unbound.

\begin{figure}[tb]
\centering
\includegraphics[angle=0,width=3.5in,draft=False]{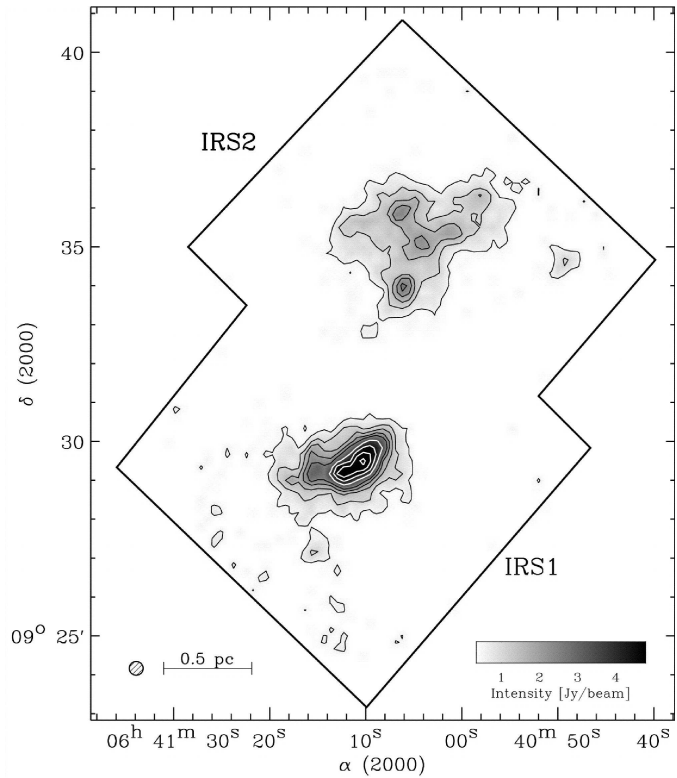}
\caption[fg11.eps]{The 870~$\mu$m continuum emission map of the IRS1 and IRS2 regions from
Williams \& Garland (2002). The intensity scale and beamsize are annotated in the lower right and
left corners of the figure, respectively. The elongated shape of IRS1 exhibits signs of substructure,
while IRS2 is more fragmented and therefore suggestive of a more evolved cluster of protostars.
\label{f11}}
\end{figure}

Nakano et al. (2003) completed high resolution H$^{13}$CO$^{+}$ (1--0) and 93 GHz continuum
observations of the IRS1 region using the Nobeyama Millimeter Array and the Nobeyama 45-meter
telescope. Four sources were identified in the resulting map, three of which were coincident
with sources identified by Ward-Thompson et al. (2000). Nakano et al. (2003) conclude that
a dense shell of gas $\sim0.12$ pc in diameter envelops IRS1, the interior of which has been
evacuated on timescales of $\sim0.1$ Myr.
Schreyer et al. (2003) examined IRS1 at high resolution using the IRAM Plateau de Bure interferometer
at 3 mm and in the CS (2--1) transition. They complement this data with 2.2, 4.6, and 11.9~$\mu$m
imaging to interpret the immediate environment around IRS1. No circumstellar disk was found around
IRS1. This young B-type star and several low-mass companions appear to be within a low-density
cavity of the remnant cloud core. The source of the large bipolar outflow is also identified as a
deeply embedded sources lying 20\arcsec\ north of IRS1. Shown in Figure 12 is the HST NICMOS image of the IRS1 field with
several embedded sources identified, including source 8, a binary which exhibits a centrosymmetric
polarization pattern consistent with circumstellar dust emission.

\begin{figure}
\centering
\includegraphics[angle=0,width=3in, draft=False]{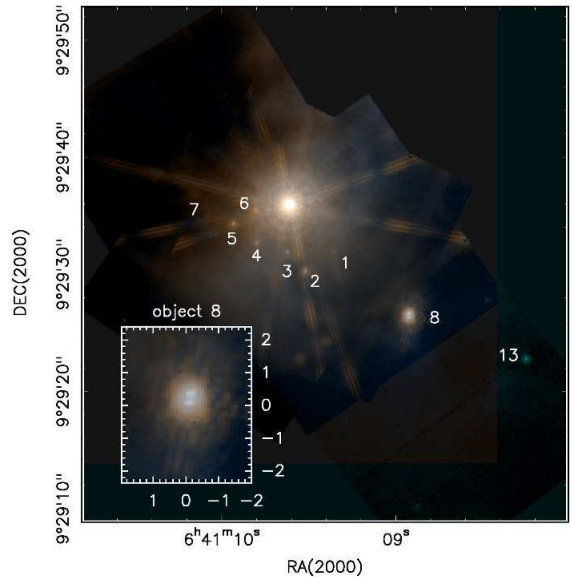}
\caption[fg12.eps]{The NICMOS composite image of the IRS1 field from Schreyer et al. (2003). The image
was created from the 1.6~$\mu$m image (blue), the mean of 1.6 and 2.2~$\mu$m images (green), and the 2.2~$\mu$m
image (red). Numerous embedded sources are evident around IRS1 including source 8, a binary which
exhibits a centrosymmetric polarization pattern consistent with dust emission.
\label{f12}}
\end{figure}

Hedden et al. (2006) used the Heinrich Hertz Telescope and the ARO 12 meter millimeter-wave telescope
to map several outflows in the northern cloud complex of the Mon OB1 association in $^{12}$CO (3--2),
$^{13}$CO (3--2), $^{12}$CO (1--0), and 870~$\mu$m continuum. Several continuum emission cores were identified
and the SEDs of these sources were constructed to derive their column densities, masses, luminosities,
and temperatures. Hedden et al. (2006) conclude that the molecular cloud complexes are maintaining their
integrity except along the axes of outflows. The outflows are found to deposit most of their energy
outside of the cloud, leading to a weak correlation between outflow kinetic energy and turbulent energy
within the clouds.

Peretto et al. (2006) surveyed the two massive cloud cores associated with IRS1 and IRS2 in dust
continuum and molecular line emission, finding 12 and 15 compact millimeter continuum sources within
each core, respectively. The millimeter sources have typical diameters of $\sim$0.04 pc and range
in mass from $\sim$2--41 M$_{\odot}$. Although similar in size to cores within the $\rho$ Oph star
forming region, the millimeter sources in NGC\,2264 exhibit velocity dispersions two to five times
greater than those of the $\rho$ Oph main cloud and the isolated cores in the Taurus-Auriga complex.
As many as 70\% of the sources within the IRS1 cloud core host Class 0 protostars that are associated with
jets of shocked H$_{2}$. In the IRS2 cloud, only 25\% of the millimeter sources are associated with
2MASS or mid-infrared point sources. Peretto et al. (2006) also find evidence for widespread infall
within both cloud cores and suggest that the IRS1 core is collapsing along its long axis
in a free-fall timescale of $\sim$1.7$\times10^{5}$ years. This is consistent with the findings
of Williams \& Garland (2002). Within this core, Peretto et al. (2006) conclude that a high-mass
(10--20 M$_{\odot}$) protostar is currently forming.

Reipurth et al. (2004c) completed a 3.6 cm radio continuum survey of young outflow sources including
NGC\,2264 IRS1 using the Very Large Array in its A configuration. Their map reveals 8 sources in the
general region of IRS1, including a source that is coincident with IRS1 itself. The sources VLA2 (MMS4)
and VLA7 appear extended and show significant collimation. Three arcminutes southeast of IRS1, Reipurth et al. (2004c)
find a bipolar radio jet with a 3.6 cm flux density of $\sim$13 mJy. The eastern portion of this jet
is comprised of at least 8 well-resolved knots that appear flattened perpendicular to the axis of the flow.
The extent of this well-collimated jet is estimated to be 28\arcsec. The western lobe of the outflow
exhibits only one large clump. Deep optical and $K-$band imaging of the central region of the jet reveal
no evidence for a source, suggesting that it may be extragalactic in origin. Trejo \& Rodr\'\i guez (2008)
compare 3.6 cm observations of this non-thermal radio jet obtained in 2006 with archived data from 1995
and find no evidence for proper motion or polarization changes. Flux density variations were found in
one knot, which is tentatively identified as the core of a quasar or radio galaxy.

Teixeira et al. (2007) present high angular resolution (1\arcsec) 1.3 mm continuum observations of the
core D-MM1 in the Spokes cluster obtained using the Submillimeter Array (SMA). They find a dense cluster of 7
Class 0 objects within a 20\arcsec$\times$20\arcsec\ region with masses ranging from 0.4 to 1.2 M$_{\odot}$.
Teixeira et al. (2007) conclude that the 1.3 mm continuum emission arises from the envelopes of the Class 0
sources, which are found to be $\sim$600 AU in diameter. The sources within the D-MM1 cluster
have projected separations that are consistent with hierarchical fragmentation.

\section{Infrared Surveys of NGC\,2264}

 One of the earliest infrared surveys of the cluster by Allen (1972) identified several sources
that were correlated with known stars, but a single bright source was identified near the tip of the
Cone Nebula that lacked an optical counterpart. This source, now referred to as Allen's infrared source
or NGC\,2264 IRS1, is recognized to be an embedded early-type (B spectral class) star.
Allen (1972) suggested that this source was the most massive and luminous star within
the cluster, a claim that has since been refuted by subsequent infrared observations.
Harvey et al. (1977) surveyed NGC\,2264 and NGC\,2244 in the mid- and far-infrared (53--175~$\mu$m)
using the Kuiper Airborne Observatory. IRS1 remained unresolved in the far infrared,
but possessed relatively cool (53--175~$\mu$m) colors.
From their luminosity estimate for IRS1 of 10$^{3}$ $L_{\odot}$, Harvey et al. (1977) concluded
that relative to compact HII regions believed to be the progenitors of massive O-type stars,
IRS1 was significantly less luminous, implying that it was an embedded intermediate-mass star of 5--10 M$_{\odot}$.
Sargent et al. (1984) completed a balloon-borne, large-scale mapping of NGC\,2264 at 70 and 130~$\mu$m,
identifying a number of far infrared sources within the cluster. The luminosity of IRS1 was found to be
3.8$\times$10$^{3}$ $L_{\odot}$, which was consistent with the earlier higher resolution observations
of Harvey et al. (1977).

Warner et al. (1977) obtained $JHKL-$ band observations of 66 members of NGC\,2264 using an
indium antimonide detector on the Mount Lemmon 1.5 meter telescope. They confirmed the existence of
infrared excesses for a significant fraction (30\%) of stars with spectral types later than A0.
Margulis et al. (1989) identified 30 discrete IRAS sources in the Mon OB1 molecular cloud complex,
18 of which were found to have Class I spectral energy distributions. From the large population of
Class I sources, Margulis et al. (1989) concluded that active star formation is still taking place within
the molecular cloud complex hosting the cluster. Neri et al. (1993) presented $uvby\beta$ and $JHKLM-$
band photometry for $\sim 50$ peculiar stars within the cluster, from which they determined a revised distance
estimate and mean extinction value. The authors also examined optical and infrared variability among the sample
stars and derive effective temperatures and log g values for each. They found no evidence for A or B-type
stars with infrared fluxes lower than expected for their observed optical magnitudes. Such stars
had been reported previously in the Ori I OB association.

The first extensive NIR imaging surveys of NGC\,2264
were completed by Pich\'{e} (1992, 1993) and Lada et al. (1993), who mapped most of the cluster
region in the $JHK-$bands. Lada et al. (1993) detected over 1650 $K-$band sources in their survey
and concluded that 360$\pm$130 were probable cluster members. Of these, 50$\pm$20\% possessed
infrared excess emission, possibly implying the presence of circumstellar disks.
Rebull et al. (2002) undertook an optical and NIR survey of the cluster, presenting photometry for over
5600 stars and spectral types for another 400.  Three criteria were used to identify circumstellar disk
candidates within the cluster: excess ($U-V$) emission, excess NIR emission ($I-K$ and $H-K$), and large H$\alpha$
equivalent widths, if spectra were available. Rebull et al. (2002) established an inner disk fraction ranging from
21\% to 56\%. No statistically significant variation was found in the disk fraction as a function of age,
mass, $I-$band mag, or ($V-I$) color.  Mass accretion rates were derived from $U-$band excesses with typical
values on the order of 10$^{-8}$ M$_{\odot}$ yr$^{-1}$.

Wang et al. (2002) completed narrowband H$_{2}$, $\nu = 1-0$ S(1) imaging of the IRS1 region and
identified at least four highly collimated jets of emission as well as several isolated knots
of H$_{2}$ emission. Some of the jets are associated with millimeter and submillimeter
sources identified by Ward-Thompson et al. (2000). Aspin \& Reipurth (2003) imaged the IRS2 region
in the NIR ($JHK$) and thermal $L'$ and $M'-$bands, finding two stars with spectra similar
to those characteristic of FU Ori type stars. The stars form a close (2\farcs8) binary and exhibit arcuate
reflection nebulae. Aspin \& Reipurth (2003) compile the ($JHKL'M'$) magnitudes for $\sim$32 stars
in the IRS2 region.

{\it Spitzer} Space Telescope is revolutionizing our understanding of the star formation process and
circumstellar disk evolution. Three-color InfraRed Array Camera (IRAC) and Multiband
Imaging Photometer for {\it Spitzer} (MIPS) images of NGC\,2264 have been released that
reveal significant structure within the molecular cloud cores as well as embedded clusters
of Class~I sources. Teixeira et al. (2006) identify primordial filamentary substructure
within one cluster such that the protostars are uniformly spaced along cores of molecular
gas in a semi-linear fashion and at intervals consistent with the Jeans length.
Shown in Figure 13 is the three-color composite image of the embedded
``Spokes'' cluster from Teixeira et al. (2006) constructed from IRAC 3.6~$\mu$m (blue), IRAC 8.0~$\mu$m (green),
and MIPS 24.0~$\mu$m (red) images of the region. Several quasi-linear structures appear to be coincident
with dust emission from dense cores of molecular gas traced at 850~$\mu$m with SCUBA by Wolf-Chase et al. (2003).
Figure 14, obtained from Teixeira et al. (2006), compares the spatial distribution of the dust emission cores
with 24~$\mu$m point sources. The sizes of the stars representing 24~$\mu$m sources are proportional to their
magnitudes.

\begin{figure}
\centering
\includegraphics[angle=0,width=\linewidth, draft=False]{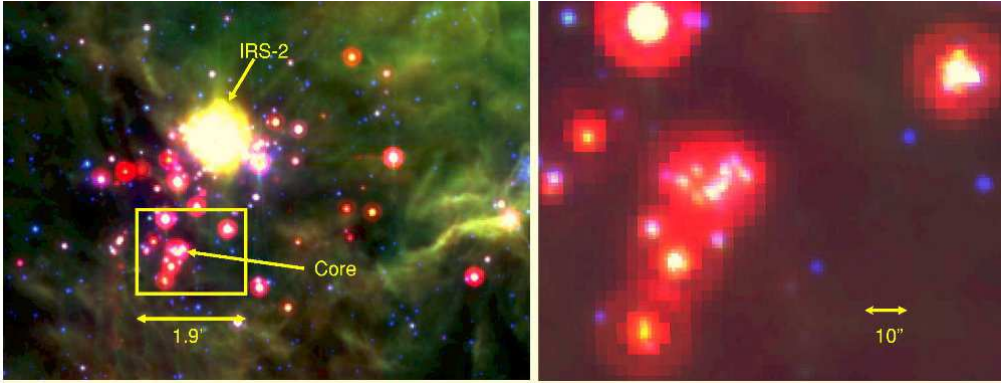}
\caption[fg13.eps]{A mid-infrared composite image of the star forming core near IRS 2 from Teixeira et al. (2006).
{\it Spitzer} IRAC 3.6 and 8.0~$\mu$m data as well as MIPS 24~$\mu$m band imaging were used to create the image.
\label{f13}}
\end{figure}

When complete, the {\it Spitzer} IRAC and MIPS surveys of NGC\,2264 will unambiguously identify the
disk-bearing population of the cluster and provide tentative characterizations of disk structure based upon the
stellar optical to mid-infrared SEDs. When merged with extant rotational period data, the {\it Spitzer}
results may resolve long-standing questions regarding the impact of circumstellar disks upon stellar rotation
and angular momentum evolution. Population statistics for the embedded clusters will also add to the stellar
census of NGC\,2264 and the greater Mon OB1 association.

\begin{figure}
\centering
\includegraphics[angle=0,width=4in, draft=False]{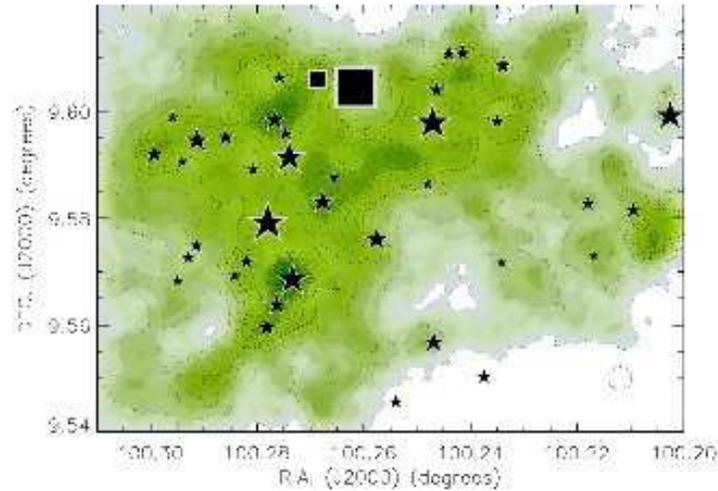}
\caption[fg14.eps]{From Teixeira et al. (2006), 850~$\mu$m dust emission from Wolf-Chase et al. (2003) with 24~$\mu$m
point sources superposed as five-pointed stars. The squares mark the positions of the two brightest 24 $\mu$m sources.
The size of the stars and squares are proportional to the magnitudes of the sources. The beam size for the 850 $\mu$m
data is indicated in the lower right. \label{f14}}
\end{figure}

\section{X-ray Surveys of NGC\,2264}

The earliest X-ray survey of NGC\,2264 was undertaken by Simon et al. (1985) with the
imaging proportional counter (IPC) aboard the {\it Einstein} Observatory. These early X-ray observations
had 1\arcmin\ spatial resolution and an energy bandwidth of 0.4--4.0 keV. All three images obtained
were centered upon S~Mon and had short integration times of 471, 1660, and 1772 s. In addition
to S~Mon, seven X-ray sources were identified by the {\it Einstein} program with X-ray luminosities
ranging from 2.4--5.2$\times$10$^{31}$ ergs s$^{-1}$. These sources were among the most X-ray
luminous of all young cluster stars observed with {\it Einstein}. ROSAT observed the cluster in 1991
March and 1992 September using the high resolution imager, HRI.  Exposure times were significantly
longer, 19.4 and 10.9 ks, respectively, resulting in the detection of 74 X-ray sources with
S/N $>$ 3.0 (Patten et al. 1994).  With the exception of A through early F-type stars, the HRI
observations detected cluster members over a range of spectral types from O7V to late-K.
Patten et al. (1994) also compared X-ray surface fluxes of solar analogs in the Pleiades,
IC 2391, and NGC\,2264, finding all three to be comparable.

Flaccomio et al. (2000) combined three archived ROSAT images of NGC\,2264 with three new
observations made approximately 15\arcmin\ southeast of the earlier epochs of data.
From these images, 169 distinct X-ray sources were identified: 133 possessed single
optical counterparts, 30 had multiple counterparts and six had no optical counterparts.
Flaccomio et al. (2000) used optical ($BVRI$) photometry from Flaccomio et al. (1999) and Sung et al. (1997)
to construct an HR diagram of the X-ray emission population of the cluster.
Ages and masses were also derived from the models of D'Antona and Mazzitelli (1997) using the
1998 updates. Comparing the ages and masses of the X-ray population with those derived
from their earlier optical survey of the southern half of the cluster, Flaccomio et al. (2000)
concluded that the X-ray sample was representative of the entire pre-main sequence population of the cluster.
Flaccomio et al. (2000) also found X-ray luminosities of known CTTSs and WTTSs to be comparable,
implying that accretion was not a significant source of X-ray emission. Using the same data set,
but with an improved determination of X-ray luminosities and a better reference optical sample,
Flaccomio et al. (2003) did find that CTTSs have on average lower X-ray luminosities with respect
to WTTSs.

Nakano et al. (2000) used The Advanced Satellite for Cosmology and Astrophysics (ASCA) to observe
NGC\,2264 in 1998 October with the Gas-Imaging Spectrometer (GIS) and the Solid state Imaging Spectrometer (SIS).
The field center of the observation was near W157, several arcmin northwest of the Cone Nebula.
Given the moderate resolution of SIS (30\arcsec),
establishing optical or infrared counterparts of the X-ray emission sources was difficult. A dozen
X-ray sources within the cluster were identified including two Class I sources, IRS1 and IRS2,
as well as several known H$\alpha$ emitters.  Nakano et al. (2000) suggest that most
of the detected hard X-ray flux originates from intermediate mass Class I sources, similar to Allen's source.

The improved spatial resolution ($\sim$1\arcsec) and sensitivity of {\it XMM-Newton} and {\it Chandra} have revolutionized X-ray
studies of young clusters and star forming regions. Optical and NIR counterparts of most X-ray sources can now be
unambiguously identified, even in clustered regions. Simon \& Dahm (2005) used deep (42 ks) {\it XMM-Newton}
EPIC observations of the northern and southern halves of NGC\,2264 to probe sites of active star formation.
The resulting integrations revealed strong X-ray emission from three deeply embedded YSOs near IRS1 and IRS2.
The brightest X-ray source was located  11\arcsec\ southwest of Allen's source and had a X-ray
luminosity of 10$^{33}$ ergs s$^{-1}$ and a temperature of 100 MK. Follow-up 1--4~$\mu$m, moderate-resolution
spectra of the sources revealed deep water ice absorption bands at 3.1~$\mu$m as well as many emission
and absorption features of HI, CO, and various metals. The NIR images of the IRS1 and IRS2 regions with
the {\it XMM} contours superposed are shown in Figure 15. Within the {\it XMM} EPIC frames, over 316
confirmed X-ray sources were identified, 300 of which have optical or NIR counterparts.
Dahm et al. (2007) find that most of these X-ray sources lie on or above the 3 Myr isochrone
of Siess et al. (2000). Given the estimated low-mass population of NGC\,2264 from variability studies and
H$\alpha$ emission surveys, the {\it XMM} sample represents only the most X-ray luminous members of the cluster.

\begin{figure}
\centering
\includegraphics[width=0.8\textwidth,angle=0, draft=False]{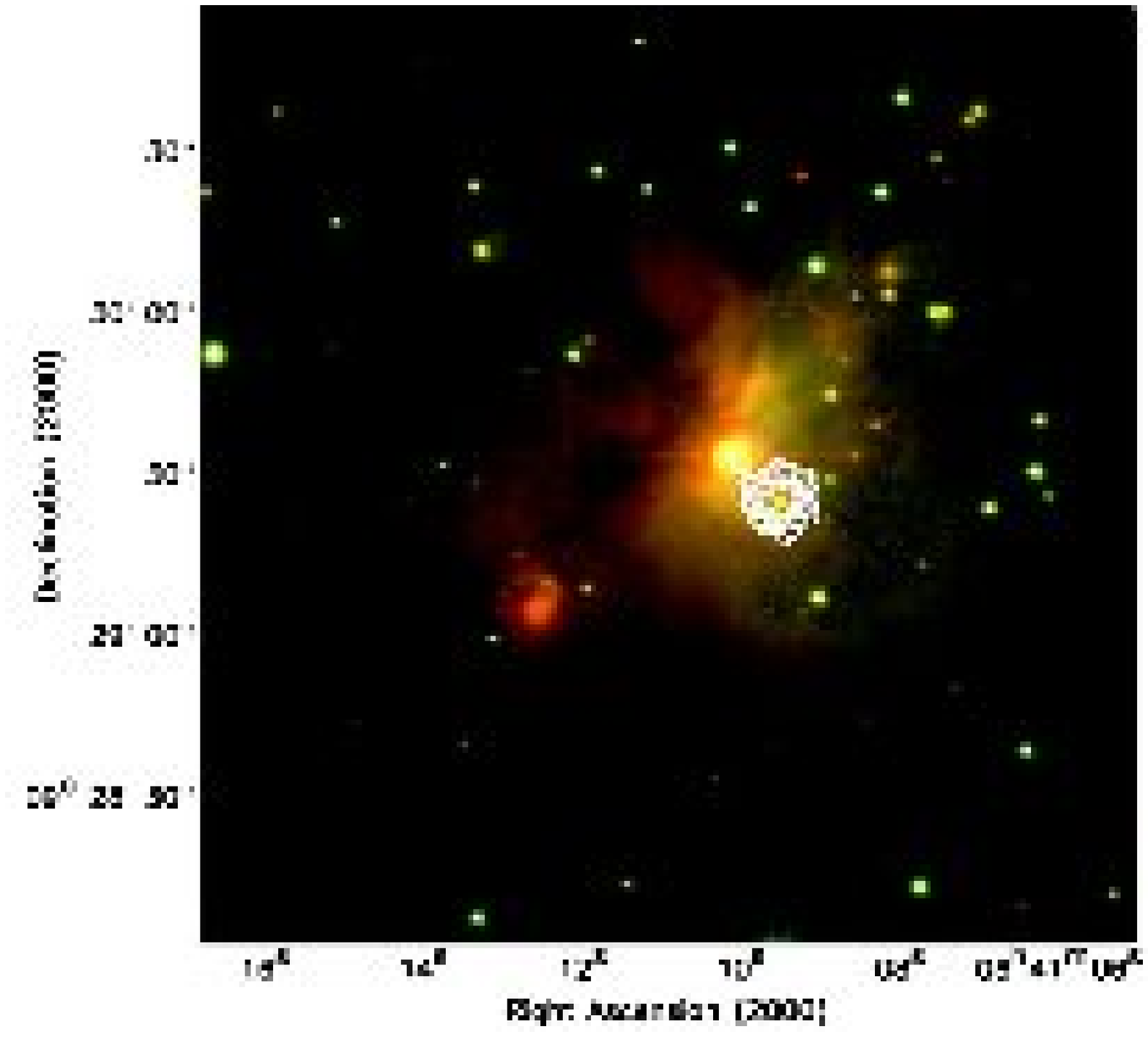}
\vspace{1.0cm}
\centering
\includegraphics[width=\textwidth,angle=0, draft=False]{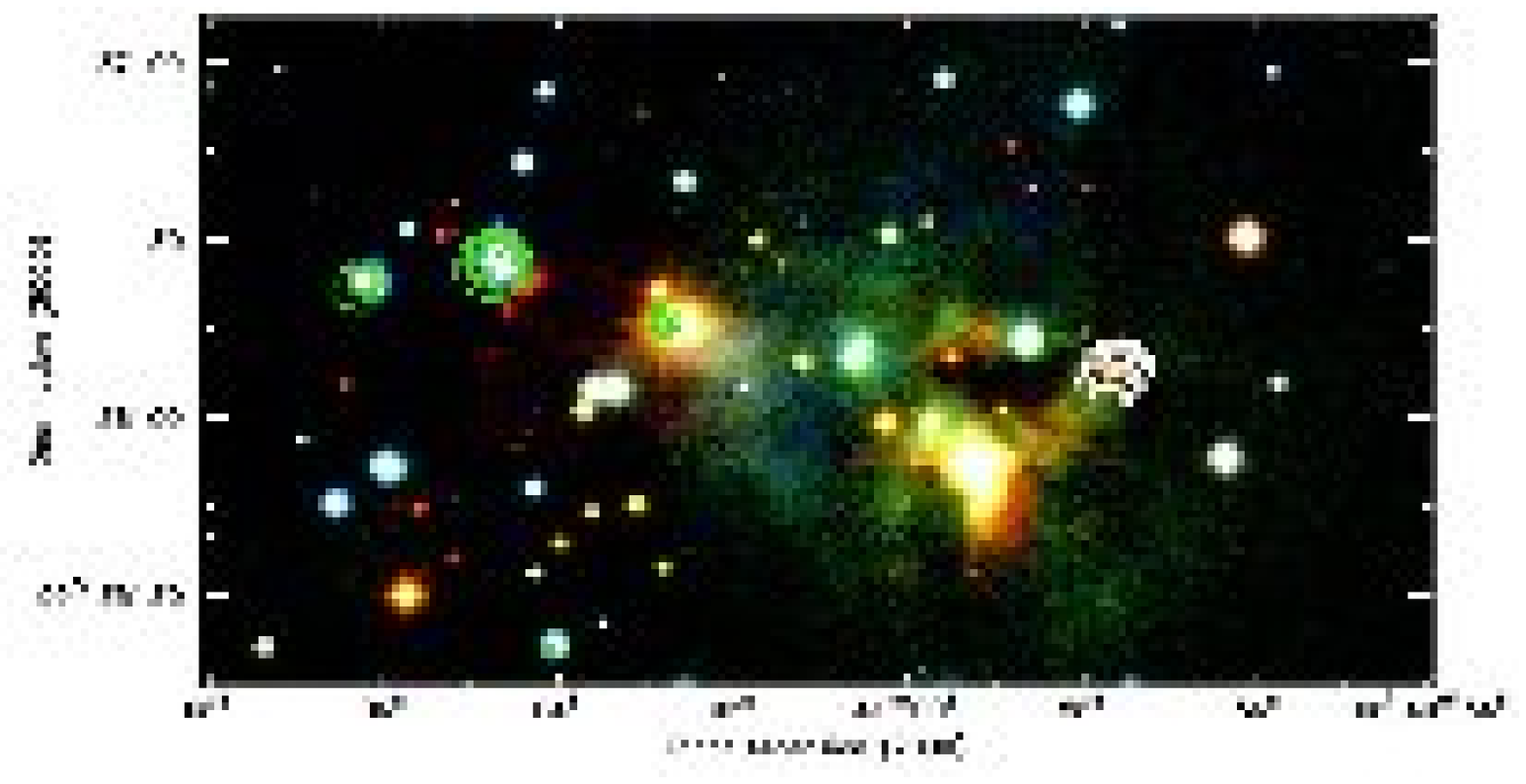}
\caption[fg15a.eps]{ ({\it a}) Near infrared ($JHK$) false color images of the embedded X-ray sources near IRS1.
Contours for the brightest X-ray emission levels are overlaid. The strongest X-ray source in the cluster is EXS-1, which is
identifiable with the infrared source 2MASS J06410954+0929250, which lies 11\arcsec\ southwest of Allen's source.
({\it b}) In the IRS2 field, the X-ray sources are from left to right: W164, W159, RNO-E (EXS-26), and 2MASS J06405767+0936082 (EXS-10).
X-ray contours are superposed in green and white. \label{f15}}
\end{figure}

The Advanced CCD Imaging Spectrometer (ACIS) onboard the {\it Chandra} X-ray Observatory was used
by Ramirez et al. (2004) to observe the northern half of NGC 2264 in 2002 February.  The field
of view of the imaging array (ACIS-I) is approximately 17\arcmin$\times$17\arcmin\ and the total
integration time of the observation was 48.1 ks. The pipeline reduction package detected 313 sources,
50 of which were rejected as cosmic ray artifacts or duplicate detections, leaving 263 probable
cluster members. Of these sources, 41 exhibited flux variability and 14 were consistent with
flaring sources (rapid rise followed by slow decay of X-ray flux). Of the confirmed sources,
213 were identified with optical or NIR counterparts.

The deepest X-ray survey of NGC\,2264 to date is that of Flaccomio et al. (2006), who obtained a
97 ks {\it Chandra} ACIS-I integration of the southern half of the cluster in 2002 October.
Within the field of view of ACIS, 420 X-ray sources were detected, 85\% of which have optical and
NIR counterparts. Flaccomio et al. (2006) found that the median fractional X-ray luminosity,
$L_{X}$/$L_{bol}$, of the sample is slightly less than 10$^{-3}$. CTTSs were found to exhibit
higher levels of X-ray variability relative to WTTSs, which was attributed to the stochastic nature
of accretion processes. Flaccomio et al. (2006) also found that CTTSs for a given stellar mass
exhibit lower activity levels than WTTSs, possibly because accretion modifies magnetic field
geometry resulting in mass loading of field lines and thus damping the heating of plasma to X-ray
temperatures (Preibisch et al. 2005; Flaccomio et al. 2006). Flaccomio et al. (2006) also find,
however, that the plasma temperatures of CTTSs are on average higher than their WTTS counterparts.

Rebull et al. (2006) combine both of these {\it Chandra} observations of NGC\,2264 in order to examine
the X-ray properties of the full cluster population. They find that the level of X-ray emission is strongly
correlated with internal stellar structure, as evidenced by an order of magnitude drop in X-ray flux among
1--2 M$_{\odot}$ stars as they turn onto their radiative tracks. Among the sample of X-ray detected stars with
established rotation periods, Rebull et al. (2006) find no correlation between $L_{X}$ and rotation rate.
They also find no statistically significant correlation between the level of X-ray flux and the presence
or absence of circumstellar accretion disks or accretion rates as determined by ultraviolet excess.

\section{Herbig-Haro Objects and Outflows in the NGC\,2264 Region}

Herbig-Haro (HH) objects and outflows are regarded as indicators of recent star formation
activity. Early surveys of the NGC\,2264 region by Herbig (1974) identified several
candidate HH objects. Adams et al. (1979) conducted a narrow-band H$\alpha$ emission survey
of the cluster and identified 5 additional HH condensations that they conclude are associated
with the molecular cloud complex behind the stellar cluster. Walsh et al. (1992) discovered
two additional HH objects in NGC\,2264 (HH 124 and HH 125) using narrow-band imaging and
low-dispersion spectra. HH 124 lies north of the cluster region and emanates from the cometary
cloud core BRC 25, which contains IRAS 06382+1017. HH 124 is composed of at least 6 knots
of emission with the western condensations exhibiting negative high velocity wings and the
eastern components positive (up to +100 km s$^{-1}$). From a large-scale CO (3--2) map of BRC 25,
Reipurth et al. (2004a) report a significant molecular outflow along the axis of HH 124 with
no source identified. Ogura (1995) identified a giant (1 pc) bow-shock
structure associated with HH 124 using narrowband ([S II] $\lambda$$\lambda$6717, 6731) CCD
imaging and slit spectroscopy. The large projected distance of the shock from its proposed source
implies a dynamical age in excess of 10,000 yrs.

\begin{figure}[!htb]
\centering
\includegraphics[width=10cm,angle=0, draft=False]{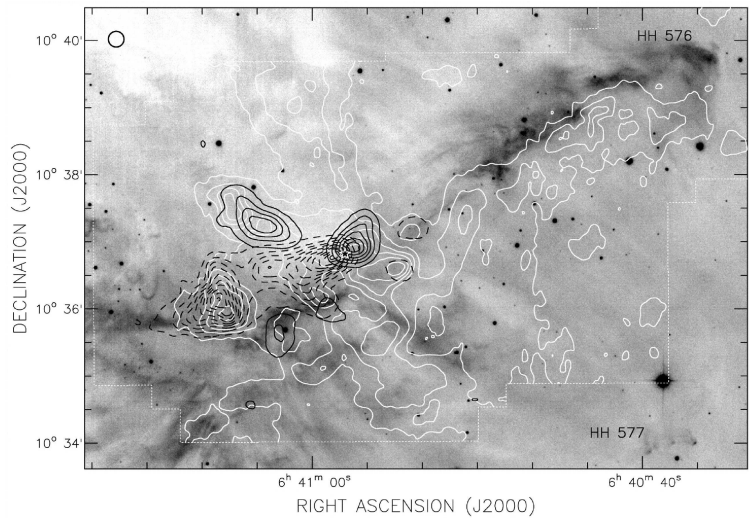}
\caption[fg16.eps]{Narrowband H$\alpha$ image of the HH 576 and 577 region with the CO (3--2)
emission contours superposed from Reipurth et al. (2004a). The extent of the region mapped in
CO is shown by the light dashed white line. Solid black contours represent blueshifted (-3 to
+3 km s$^{-1}$) CO emission, and dashed black contours redshifted (+13 to +19 km s$^{-1}$)
emission.  \label{f16}}
\end{figure}

Walsh et al. (1992) found that HH 125 is composed of at least 16 knots of emission and lies near
other known HH objects identifed previously by Adams et al. (1979). The angular extent of the HH
object implies a projected linear dimension of $\sim$0.78 pc. Walsh et al. (1992) suggested that
IRAS 06382+0945 is the exciting source for HH 125.

Wang et al. (2003) completed an extensive 2$^{\circ}$ wide-field [S II] narrowband imaging survey
of the Mon OB1 region. In the northern part of the molecular cloud, two new HH objects were discovered
(HH 572 and HH 575). Reipurth et al. (2004a) identify 15 additional HH objects in the region, some
of which have parsec-scale dimensions. One these is the giant ($\sim$5.2 pc) bipolar flow HH 571/572
which also originates from a source within BRC 25, possibly IRAS 06382+1017. The CO map of
Reipurth et al. (2004a) revealed two large molecular outflows with position angles similar to those
of HH 576 and 577, suggesting a physical association. Figure 16 shows the CO contours overlaying an
H$\alpha$ image of HH 576 and 577. The southwestern quadrant of the BRC 25 cloud core also shows
optical features that suggest significant outflow activity, but no CO counterpart was identified.
Reipurth et al. (2004a) suggested that HH 125, 225, and 226 form a single giant outflow, possibly
originating in the IRS2 region. Table 5 summarizes known HH objects in the NGC\,2264 region.

\begin{table}[tb]
\begin{small}
\begin{center}
\begin{tabular}[c]{ccccc}
\multicolumn{5}{c}{Table 5} \\
\multicolumn{5}{c}{Herbig-Haro Objects in the Mon OB1 Region} \\
\tableline
Identifier & RA (J2000) & $\delta$ (J2000) & Notes & Ref\tablenotemark{a}\\
\tableline
HH 39	 &  06 39 07    &  +08 51.9    &  NGC\,2261          & H74 \\
HH 575A &  06 40 31.6  &  +10 07 56   &   Brightest knot  & R04 \\
HH 576   &  06 40 35.9  &  +10 39 48   &  Bow shock (west)             & R04 \\
HH 577   &  06 40 36.6  &  +10 34 02   &  Brightest knot      & R04 \\
HH 571   &  06 40 46.5  &  +10 05 15   &  Brightest point        & R04 \\
HH 580   &  06 40 56.7  &  +09 32 52   &  Tip of bow                         & R04 \\
HH 582   &  06 40 56.9  &  +09 31 20   &  Western knot                   & R04 \\
HH 581   &  06 41 00.5  &  +09 32 56   &  Southern knot         & R04 \\
HH 573A &  06 41 01.9  &  +10 14 51   &  Diffuse knot in BRC 25 & R04 \\
HH 226  & 06 41 02.2   &  +09 39 49   &                                      & W03 \\
HH 124  &   06 41 02.7  &  +10 15 03   &  Bow shock     & O95 \\
HH 225  &   06 41 02.7  &  +09 44 16   &                                     & W03 \\
HH 125   &  06 41 02.8  &  +09 46 07   &                                     & W92 \\
HH 583   &  06 41 06.5  &  +09 33 16   &  Middle knot                        & R04 \\
HH 574   &  06 41 07.9	&  +10 16 19   &  Brightest knot                     & R04 \\
HH 578	 &  06 41 10.6	&  +10 20 50   &  Star at end of jet                 & R04 \\
HH 579	 &  06 41 14.4	&  +09 31 10   &  Tip of bow                         & R04 \\
HH 585   &  06 41 25.3  &  +09 24 01   &  Middle of bow                      & R04 \\
HH 572   &  06 41 28.8  &  +10 23 45   &  Eastern knot in bow      & R04 \\
HH 584	 &  06 41 38.8	&  +09 28 28   &  Central knot       & R04 \\
\tableline
\noalign{\smallskip}
\multicolumn{5}{l}{$^{a}$ H74 - Herbig (1974); A79 - Adams et al. (1979); W92 - Walsh et al. (1992);}\\
\multicolumn{5}{l}{~~~ O95 - Ogura (1995); W03 - Wang et al. (2003); R04 - Reipurth et al. (2004a)}\\
\end{tabular}

\end{center}
\end{small}
\end{table}

\section{Individual Objects of Interest in the NGC\,2264 Region}

{\bf IRS1 (Allen's Infrared Source):} First discovered by Allen (1972) in his near infrared survey of NGC\,2264, IRS1 (IRAS 06384+0932) is now recognized as a
deeply embedded, early-type (B2--B5) star lying within a massive molecular cloud core. Critical investigations
into the nature of IRS1 include those of Allen (1972), Thompson \& Tokunaga (1978), Schwartz et al. (1985),
Schreyer et al. (1997, 2003), and Thompson et al. (1997). At least one molecular outflow is associated with
IRS1 (Schreyer et al. 1997; Wolf-Chase \& Gregersen 1997) as well as a jet-like structure detected in the near
infrared. HST NICMOS imaging has revealed several point sources surrounding IRS1 assessed as solar-mass,
pre-main sequence stars (Thompson et al. 1997). The millimeter continuum and molecular line observations and mid-infrared
imaging of Schreyer et al. (2003) suggest that IRS1 is not associated with a circumstellar disk of primordial
gas and dust.
\\

\noindent {\bf W90 (LH$\alpha$25):} No discussion of NGC\,2264 would be complete without addressing the remarkable Herbig AeBe star, W90 (LH$\alpha$ 25). Herbig (1954)
lists the star as an early-A spectral type with bright H$\alpha$ emission and possible evidence for photometric variation ($\sim$0.1 mag)
relative to earlier brightness estimates by Trumpler (1930). He further noted that for a normal A2--A3 type star, LH$\alpha$ 25 is three mag fainter
than expected for the adopted distance of the cluster. Walker (1956) found the star to lie well below the ZAMS of NGC\,2264.
He further states that the star appears unreddened, but that structure within the Balmer lines and the presence of several
[Fe II] lines in its spectrum argue for a weak shell enveloping the star. Herbig (1960) revised the spectral type of
W90 to B8pe + shell, but noted that the Balmer line wings are weak relative to other late B-type stars. Poveda (1965) suggested
that stars below the ZAMS similar to W90 are surrounded by optically thick dust and gas shells, which induce neutral extinction.
Strom et al. (1971) estimated the surface gravity of W90 using its Balmer line profiles, finding log g $\sim 3$, consistent
with giant atmospheres. They further speculated that W90 is surrounded by a dust shell that absorbs 95\% of the radiated visible
light. Strom et al. (1972) confirmed the presence of dust around W90, finding an extraordinary infrared ($V-K$) excess of 3 mag.
Rydgren \& Vrba (1987) presented an SED for the star spanning from 0.35--20~$\mu$m, and concluded that W90 is observed through
an edge-on dust disk. Dahm \& Simon (2005) presented a NIR spectrum of the star (0.85--2.4~$\mu$m), which reveals strong Brackett
and Paschen series line emission as well as He I and Fe II emission. W90 remains something of an enigma, but is a
key representative of the Herbig AeBe population.
\\

\noindent {\bf KH-15D (V582 Mon):} This deeply eclipsing ($\sim$3.5 mag in $I$) K7 TTS lies just north of the Cone Nebula and the B2 star HD 47887.
The eclipse profile of KH-15D (Kearns \& Herbst 1998) suggests that an inclined knife-edge screen is
periodically occulting the star (Herbst et al. 2002). During eclipse, Herbst et al. (2002)
find that the color of the star is bluer than when outside of eclipse. These observations
and the polarization measurements of Agol et al. (2004)
support the conclusion that the flux received during eclipse is scattered by large dust grains within the
obscuring screen (Hamilton et al. 2005). By combining archived observations with recent photometry,
Hamilton et al. (2005) were able to analyze a 9-year baseline of eclipse data. Their finding suggest
that the eclipse is evolving rapidly, with its duration lengthening at a rate of 2 days per year.
Johnson et al. (2004) present results of an high resolution spectroscopic monitoring program for KH-15D
and find it to be a single-line spectroscopic binary with a period of 48.38 days, identical to the
photometric period. They conclude that the periodic dimming of KH-15D is caused by the binary motion
that moves the visible stellar component above and behind the edge of an obscuring cloud. Estimates
for the eccentricity and mass function are given. Johnson et al. (2005) present historical $UBVRI$
photometric observations of KH-15D obtained between 1954 and 1997 from multiple observatories. They
find that the system has been variable at the level of 1 mag since at least 1965. No evidence for
color variation is found, and Johnson et al. (2005) conclude that KH-15D is being occulted by an
inclined, precessing, circumbinary ring. Winn et al. (2006) use radial velocity measurements, CCD
and photographic photometry obtained over the past 50 years to examine whether a model
of KH-15D that incorporates a circumbinary disk can successfully account for its observed flux variations.
After making some refinements such as the inclusion of disk scattering, they find that the model is
successful in reproducing the observed eclipses.
\\

\begin{figure}[htb]
\centering
\includegraphics[width=12cm,angle=0, draft=False]{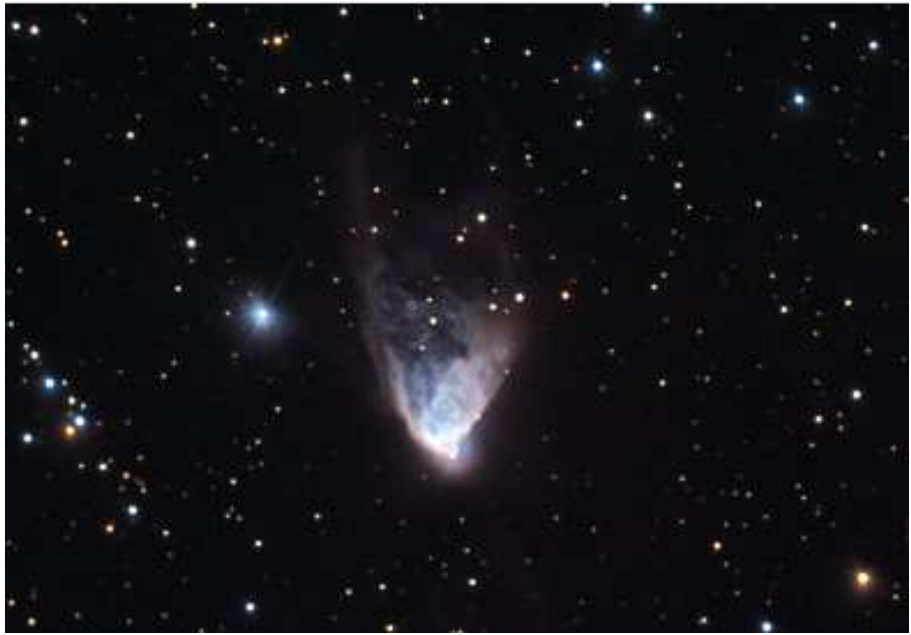}
\caption[fg17.eps]{Composite image of R Mon and NGC\,2261 obtained by Carole Westphal and Adam Block of NOAO.
\label{f17}}
\end{figure}

\noindent {\bf NGC\,2261 (Hubble's Variable Nebula):}
Over a degree south of S~Mon lies the small reflection nebula NGC\,2261, first noted by Friedrich Wilhelm Herschel
in his catalog of nebulae and stellar clusters. Shown in Figure 17 is a composite image of this object that reveals
some of the extraordinary detail associated with the nebulosity.
Hubble (1916) describes NGC\,2261 as a cometary nebula in ``the form of an equilateral triangle with a
sharp stellar nucleus at the extreme southern point." What drew Hubble's attention to the object, however,
were indisputable changes in the outline and structure of the nebula that occurred between March 1908 and
March 1916. Within the nucleus of NGC\,2261 is the irregular variable and Herbig AeBe star, R Mon, which ranges in brightness
from $V\sim$9.5 to 13 mag. The changes within the nebula, however, did not appear to coincide with the brightness variations
of R Mon. Slipher (1939) obtained a spectrum of NGC\,2261 that revealed bright hydrogen emission lines
superposed upon a faint continuum. The nova-like spectrum of the nebula was identical to that of R Mon,
even in its outlying regions. Lampland (1948) examined several hundred photographic plates of NGC\,2261
obtained over more than two decades and concluded that changes in the nebula were caused by varying degrees of veiling
or obscuration, not physical motion. Several polarization studies of the nebula have been made including
those of Hall (1964), Kemp et al. (1972), Aspin et al. (1985), Menard et al. (1988), and Close et al. (1997).
Spectroscopic studies of NGC\,2261 include those of Slipher (1939), Herbig (1968), and Stockton et al. (1975).
Herbig (1968) discovered the presence of HH 39 lying 8\arcmin\ north of the apex of NGC\,2261.
CO (1--0) observations of NGC\,2261 by Cant\'o et al. (1981) identified an elongated molecular cloud centered
upon R Mon that is interpreted as being disk-shaped in structure. Stellar winds from R Mon are proposed to
have created a bipolar cavity within the cloud, the northern lobe of which is the visible nebulosity of NGC\,2261.
Brugel et al. (1984) identify both components of the highly collimated bipolar outflow associated with R Mon.
The measured velocities for the two flows are $-74$ and $+168$ km/s, respectively. Brugel et al. (1984)
argue that the flow collimation occurs within 2000 AU of the star. Walsh \& Malin (1985) obtained deep $B$
and $R-$band CCD images of HH 39, revealing a knot of nebulosity (HH 39G) that has varied in brightness and
has a measured proper motion. A filament is also observed between HH 39 and R Mon that may be related to a
stellar wind-driven flow. Movsessian et al. (2002) determined radial velocities for several knots in the HH 39
group and find that the kinematics of the system as a whole suggest the precession of the outflow.
Polarization maps of NGC\,2261 by Aspin et al. (1985) identified small lobes close to R Mon that support the
bipolar model proposed by Cant\'o et al. (1981).
Among recent investigations of R Mon and NGC\,2261 is the adaptive optics $JHK'-$band imaging polarimetry
survey of Close et al. (1997) who find that R Mon is a close binary (0.69\arcsec\ separation). The companion
is believed to be a 1.5 M$_{\odot}$ star that dereddens to the classical T Tauri star locus. R Mon itself appears to be an unresolved point source,
but exhibits a complex of twisted filaments that extend from 1000 - 100,000 AU from the star and possibly
trace the magnetic field in the region.
Weigelt et al. (2002) use near infrared speckle interferometry to examine structure within the immediate vicinity
of R Mon at 55 mas (in $H-$band) scales. The primary (R Mon) appears marginally extended in $K-$band and significantly
extended in $H-$band. Weigelt et al. (2002) also identify a bright arc-shape feature pointing away from R Mon
in the northwesterly direction, which is interpreted as the surface of a dense structure near the circumstellar disk
surrounding R Mon. Their images confirm the presence of the twisted filaments reported by Close et al. (1997).

\begin{figure}[!p]
\centering
\includegraphics[angle=0,width=4.78in,draft=False]{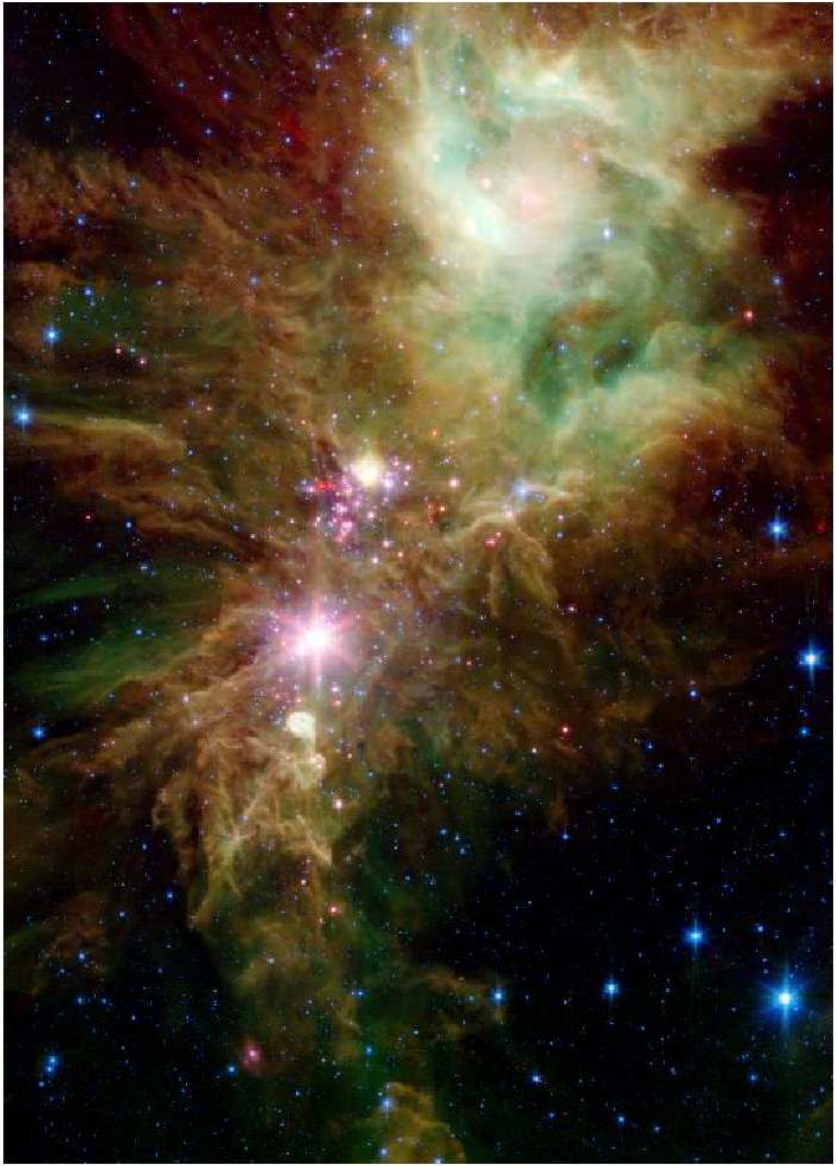}
\caption[fg18.eps]{The combined {\it Spitzer} IRAC-MIPS image of
  NGC\,2264 showing 3.6 and 4.5~$\mu$m (blue), 5.8~$\mu$m (cyan),
  8~$\mu$m (green), and 24~$\mu$m (red) emission.
\label{f18}}
\end{figure}

\section{Future Observations of NGC\,2264}

NGC\,2264 has remained a favored target for star formation studies for more than half a century,
but significant work remains unfinished. Analysis of extensive {\it Spitzer} IRAC and MIPS surveys of NGC\,2264
is nearing completion and will soon be available. The small sampling of {\it Spitzer} data
presented by Teixeira et al. (2006) of the star forming core near IRS2 provides some insight into the details
of the star formation process that will be revealed. From these datasets the disk-bearing population of the
cluster will be unambiguously identified, and tentative classifications of disk structure will be possible
by comparing observed SEDs with models. A preview of the final {\it Spitzer} image is shown in Figure 18,
which combines IRAC and MIPS data to create a composite 5-color image. The Spokes cluster is readily apparent
near image center. NGC\,2264 is perhaps best described not as a single cluster, but rather as multiple sub-clusters
in various stages of evolution spread across several parsecs. Other future observations
of the cluster that are needed include: high resolution sub/millimeter maps of all molecular cloud cores
within the region; deep optical and near infrared photometry for a complete substellar census of
the cluster; and a modern proper motion survey for membership determinations. Also of interest will
be high fidelity (HST WFPC2/WFC3) photometric studies that may be capable of reducing the inferred
age dispersion in the color-magnitude diagram of the cluster. Because of its relative proximity,
significant and well-defined stellar population, and low foreground extinction, NGC\,2264 will undoubtedly
remain a principal target for star formation and circumstellar disk evolution studies throughout the
foreseeable future.\\

{\bf Acknowledgments.}  I wish to thank the referee, Ettore Flaccomio,
and the editor, Bo Reipurth, for many helpful comments and suggestions
that significantly improved this work. I am also grateful to T. Hallas
for permission to use Figure 2, T.A. Rector and B.A. Wolpa for Figure
3, and to J.-C. Cuillandre and G. Anselmi for Figure 10, and to Carole
Westphal and Adam Block for Figure 17.  SED is supported by an NSF
Astronomy and Astrophysics Postdoctoral Fellowship under award
AST-0502381.



\end{document}